\documentclass[a4paper,12pt]{article}

\usepackage[a4paper,text={16.8cm,22.8cm}]{geometry}
\usepackage{amsmath,amssymb,bm,graphicx,color}
\usepackage{extarrows}
\usepackage[utf8]{inputenc}
\usepackage{amsthm}
\usepackage{slashed}
\usepackage{tikz}
\usepackage{rotating}
\usepackage{hyperref}
\usepackage{array}
\usepackage{color}
\usepackage{multirow}
\usepackage{bbold}
\usepackage{cite}
\usepackage{pifont}
\usepackage{tabularx}

\usepackage[export]{adjustbox}
\usepackage[normalem]{ulem}

\allowdisplaybreaks 
\renewcommand{\arraystretch}{1.2}

\begin{document}

\begin{titlepage}

\vspace*{-1.0cm}
\begin{flushright} 
P3H-21-102\\
TTP21-059 \\
IFIC/21-57

\end{flushright}

\vspace{.5 cm}
\begin{center}
\Large\bf
\boldmath
LHC Signatures of $\tau$-Flavoured Vector Leptoquarks
\unboldmath
\end{center}
\vspace{0.2cm}
\begin{center}
{\large{Jordan Bernigaud,$^{a,\,b}$ Monika Blanke,$^{a,\,b}$ Ivo de Medeiros Varzielas,$^c$ \\ Jim Talbert,$^{d, \, e}$ and Jos{\'e} Zurita$^f$}}\\
\vspace{.5cm}
{\small\sl 
${}^a$\,Institute for Astroparticle Physics (IAP), Karlsruhe Institute of Technology, Hermann-von-Helmholtz-Platz 1, D-76344 Eggenstein-Leopoldshafen, Germany,\\[0.2cm]
${}^b$\,Institute for Theoretical Particle Physics (TTP), Karlsruhe Institute of Technology, Engesserstrasse 7, D-76128 Karlsruhe, Germany\\[0.2cm]
${}^c$\,CFTP, Departamento de F\'{i}sica, Instituto Superior T\'{e}cnico, Universidade de Lisboa, Avenida Rovisco Pais 1, 1049 Lisboa, Portugal\\[0.2cm]
${}^d$\, DAMTP, University of Cambridge, Wilberforce Road, Cambridge, CB3 0WA, United Kingdom \\[0.2cm]
${}^e$\,Niels Bohr Institute, University of Copenhagen, Blegdamsvej 17, 2100 Copenhagen, Denmark\\[0.2cm]
${}^f$\,Instituto de F\'{\i}sica Corpuscular, CSIC-Universitat de Val\`encia, \\ Catedr\'{a}tico Jos\'{e} Beltr\'{a}n 2, E-46980, Paterna, Spain \\[0.2cm]
}
{\bf{E-mail}}:  jordan.bernigaud@kit.edu,  monika.blanke@kit.edu, ivo.de@udo.edu, rjt89@cam.ac.uk, jzurita@ific.uv.es
\end{center}

\vspace{0.5cm}
\begin{abstract}
\vspace{0.2cm}
\noindent 
We consider the phenomenological signatures of Simplified Models of Flavourful Leptoquarks, whose Beyond-the-Standard Model (SM) couplings to fermion generations occur via textures that are well motivated from a broad class of ultraviolet flavour models (which we briefly review). We place particular emphasis on the study of the vector leptoquark $\Delta_\mu$ with assignments $\left({\bf{3}}, {\bf{1}}, 2/3 \right)$ under the SM's gauge symmetry, $SU(3)_C \times SU(2)_L \times U(1)_Y$, which has the tantalising possibility of explaining both $\mathcal{R}_{K^{(\star)}}$ and $\mathcal{R}_{D^{(\star)}}$ anomalies. Upon performing global likelihood scans of the leptoquark's coupling parameter space, focusing in particular on models with tree-level couplings to a single charged lepton species, we then provide confidence intervals and benchmark points preferred by low(er)-energy flavour data. Finally, we use these constraints to further evaluate the (promising) Large Hadron Collider (LHC) detection prospects of pairs of $\tau$-flavoured $\Delta_\mu$, through their distinct (a)symmetric decay channels.  Namely, we consider direct third-generation leptoquark and jets plus missing-energy searches at the LHC, which we find to be complementary. Depending on the simplified model under consideration, the direct searches constrain the $\Delta_\mu$ mass up to 1500-1770 GeV when the branching fraction of $\Delta_\mu$ is entirely to third-generation quarks (but are significantly reduced with decreased branching ratios to the third generation), whereas the missing-energy searches constrain the mass up to 1150-1700 GeV  while being largely insensitive to the third-generation branching fraction.
\end{abstract}
\vfil

\end{titlepage}

%%%%%%%%%%%%%%%%%%%%%%%%%%%%%%%%%%%%%%%%%%%%%%%%%%%%%%%

\tableofcontents
\noindent \makebox[\linewidth]{\rule{16.8cm}{.4pt}}

%%%%%%%%%%%%%%%%%%%%%%%%%%%%%%%%%%%%%%%%%%%%%%%%%%%%%%%

%%%%%%%%%%%%%%%%%%%%%%%%%%%%%%%%%%%%%%%%%%%%%%%%%%%%%%%
\section{Introduction}
\label{sec:INTRO}

Despite the large amount of data collected and analysed at the Large Hadron Collider (LHC), no new Beyond-the-Standard Model (BSM) particles have been discovered yet.  Nonetheless, compelling motivations for the existence of BSM physics exist, including the unsolved electroweak (EW) hierarchy, neutrino mass, and strong CP problems, the unexplained presence of a baryon asymmetry in the Universe, the lack of a confirmed dark matter candidate, and of course the flavour puzzle. 

Besides these problems, perhaps the strongest hints for BSM physics are the deviations observed in lepton flavour universality (LFU) tests in $B$ meson decays, the so-called `flavour anomalies'.  The first indications for LFU-violating BSM interactions were found in 2012 by the BaBar collaboration \cite{Lees:2012xj,Lees:2013uzd} in the ratio 
 \begin{equation}
 \label{eq:RDstarform}
     \mathcal{R}_{D^{(\star)}} = \frac{\text{BR} \left(\overline{B} \rightarrow D^{(\star)} \, \tau^- \overline{\nu}_\tau \right)}{\text{BR} \left(\overline{B} \rightarrow D^{(\star)} \, l^- \overline{\nu}_l \right)}\,,
 \end{equation}
 with $l \in \lbrace e, \mu \rbrace$. Over the past years, measurements of the same ratios by Belle \cite{Huschle:2015rga,Hirose:2016wfn,Belle:2016ure,Belle:2016dyj,Belle:2019gij} and LHCb \cite{Aaij:2015yra,LHCb:2017smo,Aaij:2017deq} have confirmed the tension with the SM prediction, with the latest HFLAV average exhibiting a $3.4\sigma$ deviation from the SM \cite{HFLAV:2019otj}. 
 
 Due to the size of the BSM contribution required to resolve the anomaly -- an  $\mathcal{O}(10\%)$ enhancement at the matrix element level -- new physics plausibly has to enter the relevant $b\to c\tau\nu$ transition at tree level. Possible BSM scenarios then include the exchange of a new colour-singlet charged scalar (charged Higgs) \cite{Crivellin:2012ye,Crivellin:2013wna,Celis:2012dk,Ko:2012sv,Crivellin:2015hha} or vector ($W'$) boson \cite{He:2012zp,Greljo:2015mma,Boucenna:2016wpr,He:2017bft}, or of a colour-triplet scalar or vector leptoquark (LQ) \cite{Alonso:2015sja,Calibbi:2015kma,Fajfer:2015ycq,Barbieri:2015yvd,Barbieri:2016las,Hiller:2016kry,Deshpande:2012rr,Tanaka:2012nw,Sakaki:2013bfa,Freytsis:2015qca,Bauer:2015knc,Becirevic:2016yqi,Becirevic:2018afm}. The latter have the advantage of being less stringently constrained by precision EW constraints and direct LHC searches. 
 
 Of particular interest is the isospin-singlet vector LQ,\footnote{In what follows we will use the notation and nomenclature of \cite{deMedeirosVarzielas:2019lgb,Bernigaud:2019bfy,Bernigaud:2020wvn} when considering flavour structures, and \cite{Aebischer:2019mlg} for other Lagrangian parameters. We will refer to $\Delta^\mu$ as the `vector LQ singlet', which is sometimes denoted $U_1$ in the literature.}
\begin{equation}\label{eq:Delta}
\Delta^\mu \sim \left({\bf{3}}, {\bf{1}}, 2/3 \right)\,,
\end{equation}
whose respective charge assignments under the SM gauge group, $\mathcal{G}_{SM} \equiv SU(3)_C \times SU(2)_L \times U(1)_Y$, are given on the right hand side of \eqref{eq:Delta}. In contrast to scalar LQ solutions, the coupling structure of $\Delta^\mu$ is not constrained by proton decay \cite{Assad:2017iib}.
Besides that, $\Delta^\mu$ is contained in the Pati-Salam gauge group $SU(4)_C \times SU(2)_L \times SU(2)_R$ unifying quarks and leptons \cite{Pati:1974yy}, thereby providing an appealing ansatz for the construction of an ultraviolet (UV)-complete model \cite{Assad:2017iib,DiLuzio:2017vat,Calibbi:2017qbu,Bordone:2017bld,Barbieri:2017tuq,Blanke:2018sro}.
 
 From the phenomenological perspective, the vector LQ singlet gains additional appeal from the fact that it is the only single-particle solution to anomalies in both $\mathcal{R}_{D^{(\star)}}$ and $\mathcal{R}_{K^{(\star)}}$. The latter ratios, defined as
 \begin{equation}
     \mathcal{R}_{K^{(\star)}} =  \frac{\text{BR} \big({B} \rightarrow K^{(\star)} \,\mu^+\mu^- \big)}{\text{BR} \big({B} \rightarrow K^{(\star)} \, e^+ e^- \big)}\,,
 \end{equation}
exhibit a combined $\sim 4\sigma$ tension 
\cite{Alguero:2021anc,Altmannshofer:2021qrr} with the SM in LHCb \cite{LHCb:2017avl,LHCb:2021trn,LHCb:2021lvy} and Belle \cite{BELLE:2019xld,Belle:2019oag} data. The latter anomaly is further supported by the fact that deviations from the SM predictions are also showing in other observables sensitive to the quark-level $b\to s\mu^+\mu^-$ transition, such as $P'_5$, $\text{BR}(B_s\to \phi\mu^+\mu^-)$ and $\text{BR}(B_s\to\mu^+\mu^-)$. 

The new physics scale required to address the anomaly in $ \mathcal{R}_{D^{(\star)}}$ is as low as a few TeV, and therefore any underlying BSM particle(s) responsible for the (potential) new physics may be within the expected mass reach of the LHC or, eventually, future high(er)-energy colliders --- see e.g.~\cite{Allanach:2017bta,Garland:2021ghw,Asadi:2021gah,Allanach:2019zfr}.  Numerous studies of the LHC phenomenology of LQs responsible for the 
$\mathcal{R}_{D^{(\star)}}$ anomaly exist, ranging from resonant LQ pair- and single-production,  $t$-channel LQ exchange, and other non-resonant processes, see e.g.~\cite{Mandal:2015vfa,Bhaskar:2021gsy,Aydemir:2019ynb,Baker:2019sli,Aebischer:2019mlg,Iguro:2020keo,Endo:2021lhi,Cornella:2021sby,Angelescu:2018tyl,Feruglio:2018fxo,Angelescu:2021lln,Greljo:2018tzh,Husek:2021isa,Haisch:2020xjd}.  

Of these, resonant LQ production processes yield the most direct access to the parameters of the LQ model. LQ pair production is driven by QCD interactions, and hence for a given LQ representation its cross-section yields a direct determination of the LQ mass. The branching ratios of the LQ into different final states then determine the relative coupling strengths to fermions. In combination with the observed values for $\mathcal{R}_{D^{(*)}}$, which are driven by the product of the relevant LQ coupling parameters, the measurement of LQ branching ratios into the various final states then allows for a complete determination of the parameters of the simplified model. We will elaborate more on this point in Section \ref{sec:COMPLEMENTARITY}.
In this paper, after reviewing the simplified models we consider, we take specific textures of the LQ couplings to fermions that arise from specific flavour hypotheses and analyze the impact of LHC searches on the respective models.

The remainder of the paper develops as follows:  in Section \ref{sec:THEORY} we review the theory formalism embedded in our simplified models, including the symmetry motivation for studying particular LQ flavour patterns.  In Section \ref{sec:SMELLI} we use {\tt{smelli}}  \cite{Aebischer:2018iyb} to perform a global likelihood scan of said couplings, and ultimately isolate a preferred parameter space at the $1\sigma$ and $2\sigma$ confidence level.  This then provides benchmark points to study collider phenomenology in Section \ref{sec:LHC}, where we perform a reinterpretation of several LHC searches for our $\tau$-flavoured isospin-singlet vector LQ.
Finally, we provide a summary and outlook in Section \ref{sec:CONCLUDE}.

%%%%%%%%%%%%%%%%%%%%%%%%%%%%%%%%%%%%%%%%%%%%%%%%%%%%%%%
\section{Theoretical Framework}
\label{sec:THEORY}

In this Section we review the simplified model describing the dynamics of the vector LQ singlet $\Delta^\mu$ in \eqref{eq:Delta}. We start by introducing the underlying Lagrangian in Section \ref{subsec:Lagrangian}. Subsequently in Section \ref{sec:SMFL} we turn to the discussion of simple LQ coupling structures motivated by flavour symmetries. Turning our attention to one specific case, the $\tau$-isolation pattern, we then outline how the measurements of both the LFU ratios $\mathcal{R}_{D^{(*)}}$ and the LQ pair-production and decay rates at the LHC collude in the determination of the simplified model parameters.

%%%%%%%%%%%%%%%%%%%%%%%%%%%%%%%%%%%%%%%%%%%%%%%%%%%%%%%
\subsection{Simplified Model Lagrangian}
\label{subsec:Lagrangian}
When added to the SM field content, the vector LQ singlet $\Delta^\mu$ introduced in \eqref{eq:Delta} sources the following kinetic term and gauge interactions
\begin{equation}
\label{eq:DeltaLagrangian}
\mathcal{L} \supset - \frac{1}{2} \Delta^\dagger_{\mu\nu} \Delta^{\mu \nu} + i g_s (1-k_s) \Delta_{\mu}^\dagger T^A \Delta_{\nu} G^{A,\mu\nu} + i\frac{2 \,g^\prime\, (1-k_Y)}{3} \Delta_{\mu}^\dagger \Delta_{\nu} B^{\mu \nu} \,,
\end{equation}
where the LQ field strength is given by 
\begin{equation}
\label{e:fieldstrength}
\Delta_{\mu \nu} = D_\mu \Delta_{\nu} - D_\nu \Delta_\mu \, ,
\end{equation}
in terms of the gauge-covariant derivative
\begin{equation}
\label{eq:gaugederivative}
D_\mu = \partial _\mu + i g_s T^A G_\mu^A + i \frac{2g^\prime}{3} B_\mu \,.
\end{equation}
In the above equations $T^A$ are colour generators, $g_s$ and $g^\prime$ are the standard QCD and hypercharge gauge couplings, and we study two scenarios for the $k_{s,Y}$ parameters: ${\bf{(A)}}$ 
$k_Y = k_s = 0$, which  tames divergences in LQ-gauge boson scattering and dipole processes --- see \cite{Ferrara:1992yc,Barbieri:2015yvd} for details --- and is also motivated by UV-completing the vector LQ singlet as a gauge boson of an extended gauge symmetry \cite{Barbieri:2016las}, and ${\bf{(B)}}$ 
$k_Y = k_s = 1$, which corresponds to the so-called minimal coupling scenario as may appear in strongly coupled UV models \cite{Baker:2019sli}. For simplicity we will also refer to Scenarios ${\bf{(A)}}$ and ${\bf{(B)}}$ as $k=0$ and $k=1$, respectively.

In addition to the above interactions, gauge-invariant coupling terms of the form\footnote{While right-handed neutrinos play no role throughout our phenomenological analysis, we introduce them here in order to allow for a complete discussion of flavour symmetries.} 
\begin{align}
\mathcal{L} \supset x_{ij}^{LL} \bar{Q}_{L}^{i,a} \gamma^{\mu} \Delta_{\mu} L_{L}^{j,a} + x_{ij}^{RR} \bar{d}^{i}_{R} \gamma^{\mu} \Delta_{\mu} e_{R}^{j} + x_{ij}^{\overline{RR}} \bar{u}_{R}^{i} \gamma^{\mu} \Delta_{\mu} \nu_{R}^{j} + \text{h.c.}\,,
\label{eq:LLyukSU2}
\end{align}
also appear, where $\lbrace i,j \rbrace$ denote flavour indices and $a$ is an SU(2) index. For a thorough review of the physics of LQs, see e.g. \cite{Dorsner:2016wpm}.  Moving to the SM fermion mass basis via
\begin{align}
\nonumber
u_{L} &\rightarrow U_{u} u_{L}\,, \,\,\,\,\,\,\,\,\,\,\,\,\,\,\,\,\,\,\,\,\, d_{L} \rightarrow U_{d} d_{L}\,, \,\,\,\,\,\,\,\,\,\,\,\,\,\,\,\,\,\,\,\,\, l_{L}\rightarrow U_{l} l_{L}\,, \,\,\,\,\,\,\,\,\,\,\,\,\,\,\,\,\,\,\,\,\, \nu_{L} \rightarrow U_{\nu} \nu_{L} \,,  \\
\label{eq:rotate2}
u_{R} &\rightarrow U_{U} u_{R}\,,\,\,\,\,\,\,\,\,\,\,\,\,\,\,\,\,\,\,\,\, d_{R} \rightarrow U_{D} d_{R}\,, \,\,\,\,\,\,\,\,\,\,\,\,\,\,\,\, E_{R} \rightarrow U_{E} E_{R}\,,\,\,\,\,\,\,\,\,\,\,\,\,\,\,\,\,\,\,  \nu_{R} \rightarrow U_{R} \nu_{R} \, , 
\end{align}
and further decomposing SU(2)$_L$ indices, one then finds that \eqref{eq:LLyukSU2} expands to
\begin{align}
\label{eq:LLyukFLAV}
\mathcal{L} &\supset (U_{u}^{\dagger} x^{LL} U_{\nu})_{ij} \bar{u}^{ i}_{L} \gamma^{\mu} \Delta_{\mu} \nu_{L}^{j} +  (U_{d}^{\dagger} x^{LL} U_{l})_{ij} \bar{d}^{i}_{L} \gamma^{\mu} \Delta_{\mu} l_{L}^{j}  \\
\nonumber
& + (U_{D}^{\dagger}x^{RR} U_{E})_{ij} \bar{d}^{i}_{R} \gamma^{\mu} \Delta_{\mu} E_{R}^{j} + (U_{U}^{\dagger} x^{RR} U_{R})_{ij}\bar{u}^{ i}_{R} \gamma^{\mu} \Delta_{\mu} \nu_{R}^{j}  \\
\nonumber
&+ \text{h.c.}\,,
\end{align}
such that the novel BSM interactions between generations of quarks and leptons are manifest, including terms with left-left (LL) and right-right (RR) chiral structure (although we do not consider the terms with RR chiral structure in our analysis below).  As is clear, these couplings are all $3 \times 3$ matrices in flavour space, such that (e.g.)\footnote{Observe that in \cite{deMedeirosVarzielas:2019lgb,Bernigaud:2019bfy} the vector LQ singlet $d-l$ coupling features an additional superscript, $\lambda_{dl}^{V_1}$. Since we will not study the scalar triplet or vector triplet states in this work, we remove this superscript for simplicity.}
\begin{equation}
\label{eq:genericyuk}
\left(U_{d}^{\dagger}\,x^{LL}\,U_{l} \right) \equiv \lambda_{dl} =
\left(
\begin{array}{ccc}
\lambda_{de} & \lambda_{d\mu} & \lambda_{d\tau}  \\
\lambda_{se}  & \lambda_{s\mu}  & \lambda_{s\tau}     \\
\lambda_{be}  & \lambda_{b\mu}   & \lambda_{b\tau}  
\end{array}
\right) \, ,
\end{equation}
so implying that the LL $u-\nu$ coupling is related to $\lambda_{dl}$ via SU(2)$_L$ rotations, $\lambda_{u\nu} = U_\text{CKM} \, \lambda_{dl} \, U_\text{PMNS}$.  Here $U_\text{CKM}$ and $U_\text{PMNS}$ are the standard quark and lepton mixing matrices of the SM, defined by
\begin{equation}
U_\text{CKM} \equiv U_{u}^\dagger U_d\,, \,\,\,\,\,\,\,\,\,\, U_\text{PMNS} \equiv U_l^\dagger U_\nu\,,
\end{equation}
and whose matrix elements are constrained by a host of low-energy precision flavour data --- see e.g. \cite{Zyla:2020zbs,Esteban:2020cvm}.  

Finally, the RR terms of \eqref{eq:LLyukFLAV} are, a priori, fully independent. In the remainder of our analysis we will set these couplings to zero, which we motivate below.

%%%%%%%%%%%%%%%%%%%%%%%%%%%%%%%%%%%%%%%%%%%%%%%%
\subsection{Simplified Models of Flavourful Leptoquarks}
\label{sec:SMFL}
In general, one can study the phenomenology of \eqref{eq:genericyuk} with arbitrary values/shapes for the couplings $\lambda_{dl}$.  However, it is appealing to instead examine textures that are motivated by both experimental and theoretical considerations.  To that end, we will study lepton isolation patterns of the form 
\begin{equation}
\label{eq:yukeisolation}
\lambda^{[e]}_{dl} = 
\left(
\begin{array}{ccc}
{\color{red}\lambda_{de}} & 0 & 0  \\
\lambda_{se}  & 0 & 0    \\
\lambda_{be}  & 0  & 0 
\end{array}
\right), \,\,\,\,\,
\lambda^{[\mu]}_{dl} = 
\left(
\begin{array}{ccc}
0 & {\color{red}\lambda_{d\mu}} & 0  \\
0 & \lambda_{s\mu} & 0    \\
0 & \lambda_{b\mu} & 0 
\end{array}
\right), \,\,\,\,\,
\lambda^{[\tau]}_{dl} = 
\left(
\begin{array}{ccc}
0 & 0 & {\color{red}\lambda_{d\tau}}  \\
0  & 0 & \lambda_{s \tau}     \\
0  & 0  & \lambda_{b \tau}  
\end{array}
\right)   \, ,
\end{equation}
where the meaning of the red entries in the first row will be explained below.  Such matrices are some of the minimal patterns motivated by the flavour-symmetry breaking embedded in  the Simplified Models of Flavourful Leptoquarks (SMFL) developed in \cite{deMedeirosVarzielas:2019lgb,Bernigaud:2019bfy,Bernigaud:2020wvn}. The principle assumption of SMFL is that the LQ couplings to fermions of (e.g.) \eqref{eq:genericyuk} are invariant under Abelian residual family symmetries (RFS),
\begin{equation}
\label{eq:symmetryexists}
\exists \,\, \lbrace Q, L \rbrace,\,\,\,\, T^{\dagger}_{Q}\, \lambda_{QL}\,T_{L} \overset{!}{=} \lambda_{QL} \,.
\end{equation}
Here $T_{Q,L}$ are (reducible) generator representations of said RFS in arbitrary quark (Q) or lepton (L) family sectors, which simultaneously act on the SM Yukawa sector, where it is well known that each family's mass sector is invariant under $\text{U}(1)^3$ RFS (in the broken phase) and that, if present, a Majorana neutrino mass term is instead invariant under a Klein $\mathbb{Z}_2 \times \mathbb{Z}_2$ \cite{King:2013eh,Lam:2007qc}.  When RFS are interpreted as remnants of the breakdown of a UV parent symmetry, e.g. through a breaking chain  
\begin{equation}
\label{eq:GF}
\mathcal{G_{F}}  \rightarrow \begin{cases}
				\mathcal{G_{L}}   \rightarrow \begin{cases} 
										\mathcal{G_{\nu}}
										\\
										\mathcal{G_{\text{l}}}
										\end{cases} \\
				\mathcal{G_{Q}} \rightarrow \begin{cases}
										\mathcal{G_{\text{u}}}
										\\
										\mathcal{G_{\text{d}}}
										\end{cases}
				\end{cases}
\end{equation}
where $\mathcal{G}_{u,d,\nu,l}$ denote the RFS controlling infrared (IR) flavour structures and $\mathcal{G}_{\mathcal{F}, \mathcal{L}, \mathcal{Q}}$ are larger parent flavour groups,\footnote{The parent group can be continuous or discrete, Abelian or non-Abelian.} they can be used to algorithmically study the origins of CKM and PMNS mixing matrices \cite{Lam:2007qc,Hernandez:2012ra,Lam:2012ga,Holthausen:2012wt,King:2013vna,Holthausen:2013vba,Lavoura:2014kwa,Fonseca:2014koa,Joshipura:2014qaa,Talbert:2014bda,Yao:2015dwa,Lu:2016jit, Varzielas:2016zuo}, control flavour-changing neutral currents in multi-Higgs-doublet models \cite{deMedeirosVarzielas:2019dyu}, and of course structure the LQ couplings of interest here.  Critically, this analysis can be done without reference to the details of the UV flavour model's dynamics, and is therefore a largely model-independent formalism for studying (B)SM flavour.  

We leave the details of the RFS mechanism embedded in SMFL  to \cite{deMedeirosVarzielas:2019lgb,Bernigaud:2019bfy,Bernigaud:2020wvn}, and proceed by considering \eqref{eq:symmetryexists} with respect to the $d-l$ operator only,\footnote{In \cite{deMedeirosVarzielas:2019lgb} the consequences of applying \eqref{eq:symmetryexists} to all relevant family sectors were explored.  While the presence of more symmetry removes parametric degrees of freedom in a simplified model setup, it is also more challenging to accommodate in UV flavour models -- cf. e.g. \cite{Bernigaud:2019bfy,Bernigaud:2020wvn} for some discussion on this point.} where the $3 \times 3$ coupling is constrained entry-by-entry through the RFS relation 
\begin{equation}
\label{eq:LQoverconstrainV3}
    \left(
\begin{array}{ccc}
e^{i(-\alpha_{d}+\alpha_{l})}\,\lambda_{de} & e^{i(-\alpha_{d}+\beta_{l})}\, \lambda_{d\mu} &  e^{i(-\alpha_{d}+\gamma_{l})}\,\lambda_{d\tau}  \\
e^{i(-\beta_{d}+\alpha_{l})}\,\lambda_{se}  & e^{i(-\beta_{d}+\beta_{l})}\, \lambda_{s\mu}  & e^{i(-\beta_{d}+\gamma_{l})}\, \lambda_{s\tau}     \\
e^{i(-\gamma_{d}+\alpha_{l})}\,\lambda_{be}  & e^{i(-\gamma_{d}+\beta_{l})}\,\lambda_{b\mu}   &  e^{i(-\gamma_{d}+\gamma_{l})}\,\lambda_{b\tau}  
\end{array}
\right)
  \overset{!}{=}
  \left(
\begin{array}{ccc}
\lambda_{de} & \lambda_{d\mu} & \lambda_{d\tau}  \\
\lambda_{se}  & \lambda_{s\mu}  & \lambda_{s\tau}     \\
\lambda_{be}  & \lambda_{b\mu}   & \lambda_{b\tau}  
\end{array}
\right)\,.
\end{equation}
This invariance is clearly not realized in the absence of special relationships amongst the phases of the RFS generator, which are themselves IR realizations of UV flavour-symmetry breaking in specific directions of flavour space.  Indeed, the patterns of  \eqref{eq:yukeisolation} appear when one of the $T_l$ phases is also equal to $\beta_d = \gamma_d$, e.g. $\alpha_l = \beta_d = \gamma_d$ (which gives the first texture, etc.).  These matrices are given in the fermion mass basis, and the red entries in the top row of \eqref{eq:yukeisolation} highlight that it is perhaps more interesting to consider RFS which distinguish at least two fermion generations, thereby forcing $\alpha_d \neq \beta_d = \gamma_d$, which forbids these $d$-quark entries.  It was also noted in \cite{deMedeirosVarzielas:2019lgb} that zero entries in the first row are consistent with scenarios where the RFS successfully controls the dominant Cabibbo mixing observed in the CKM matrix, thereby connecting potentially anomalous signals of new physics with partial solutions to the SM's longstanding flavour puzzle.  Furthermore, it was generically shown that \eqref{eq:yukeisolation} can arise from the breakdown of non-Abelian family symmetries \cite{Bernigaud:2019bfy}, which can be described by an effective Lagrangian composed of non-renormalizable interactions between scalar flavons and SM fermion multiplets \cite{Bernigaud:2020wvn,Varzielas:2015iva}.  In short, evidence of new SMFL physics can be directly connected to more complete models of (B)SM flavour physics.

%%%%%%%%%%%%%%%%%%%%%%%%%%%%%%%%%%%%%%%%%%%%%%%%%%%%%%%%%%%
\subsection[On $\mathcal{R}_{D^{(*)}}$ and Collider Complementarity]{\boldmath On $\mathcal{R}_{D^{(*)}}$ and Collider Complementarity}
\label{sec:COMPLEMENTARITY}

Precision flavour constraints from $B$-meson decays and other low-energy processes give information that is potentially complementary to direct searches at the LHC.  Consider the LFU ratio $\mathcal{R}_{D^{(\star)}}$, which for the flavoured LQ $\Delta_\mu$ we consider is approximately given by \cite{Hiller:2016kry}
\begin{align}
    \label{eq:RDstarAPPROX}
    \mathcal{R}_{D^{(\star)}} &\simeq \mathcal{R}_{D^{(\star)}}^{\text{SM}}\cdot \left[ 1 + \frac{1}{\sqrt{2} G_F V_{cb}} \, \text{Re} \left(\lambda \lambda^\star \vert_\tau - \lambda \lambda^\star \vert_l  \right) \left(\frac{\text{TeV}}{M_\Delta} \right)^2 \right]\,,
\end{align}
including only linear matching effects to the dimension-six $(V-A) \otimes (V-A)$ operator $ \left[\overline{c} \gamma^\mu(1-\gamma_5) b \right]\left[\overline{l} \gamma_\mu(1-\gamma_5) \nu \right]$ in the weak effective theory (WET) (which  holds up to $\mathcal{O}(10 \%)$ corrections to the BSM contribution).\footnote{Note of course that the analysis in Section \ref{sec:SCANS} accounts for running effects, etc.  Here we are simply making a qualitative (and motivating) point.}  On the other hand, for a two-body decay into a given quark-lepton pair, the branching ratio (BR) for a particular vector LQ decay channel is given by (see e.g. \cite{Dorsner:2016wpm})
\begin{equation}
\label{eq:decaywidth}
   \text{BR}\left(\Delta \rightarrow QL \right) \simeq \frac{\vert \lambda_{QL} \vert^2}{\sum_{\lbrace QL \rbrace} \vert \lambda_{QL} \vert^2}\, \,\,\,\,\,\text{where} \,\,\,\,\, m_{Q,L} \rightarrow 0\,.
\end{equation}
Neglecting quark and lepton masses is an excellent approximation for the LQ mass scales we consider. 

%------------------------------------------------------------------------
\begin{figure}[tp]
\centering
  \includegraphics[width = 0.6 \textwidth]{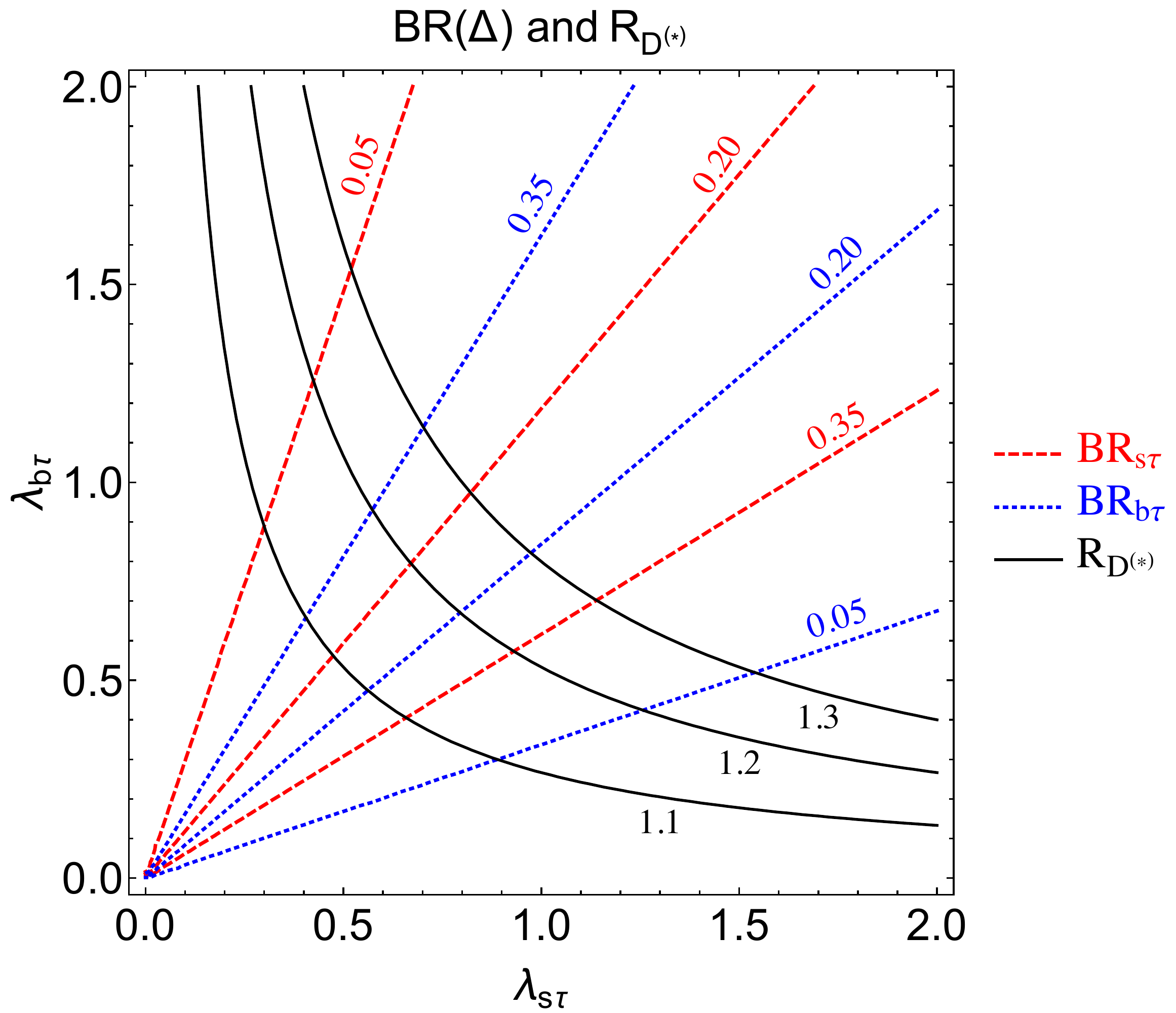}
  \caption{~\label{fig:RDLHCcomp} A comparison of contours of $\mathcal{R}_{D^{(\star)}}$, normalised to their SM values, and $\text{BR}\left(\Delta \rightarrow s\tau (b\tau) \right)$ in black and red (blue), respectively.}
\end{figure}
%------------------------------------------------------------------------

Comparing \eqref{eq:RDstarAPPROX} to \eqref{eq:decaywidth}, one sees that the former constrains a product of LQ couplings, whereas the latter constrains a ratio.  Hence, in the event of a discovery being made at the LHC, we note that combining this information would allow for a direct experimental probe at the level of individual couplings in the overall flavour matrices $\lambda_{dl, u\nu}$.  We show this qualitatively in Fig. \ref{fig:RDLHCcomp}, given the approximations in \eqref{eq:RDstarAPPROX}-\eqref{eq:decaywidth}, where we have presented experimentally relevant values of $\mathcal{R}_{D^{(\star)}}$ and various values of $\text{BR}\left(\Delta \rightarrow s\tau(b\tau)\right)$, in the context of the $\tau$-isolation model $\lambda^{[\tau]}_{dl}$.  Note that in this model SU(2) rotations lead to non-negligible contributions from the $u-\nu$ sector, cf. Fig. \ref{fig:BR_LQ_tauiso_2TEV}.

In order to fully determine the parameters of the SMFL, also the LQ mass $M_\Delta$ and the LQ coupling parameters $k_s,k_Y$ need to be measured. $k_s$ and $M_\Delta$ can be accessed through the LQ pair-production cross-section in $pp$ collisions and the invariant mass of the LQ decay products. The parameter $k_Y$ is more difficult to access, as it is responsible for the coupling strength of the LQ to the photon and, to a lesser extent, to the $Z$ boson. 

%%%%%%%%%%%%%%%%%%%%%%%%%%%%%%%%%%%%%%%%%%%%%%%%%%%%%%%
\section{Precision Constraints and Global Likelihoods}
\label{sec:SMELLI}
Our goal is to now provide a robust examination of the SMFL parameter space favoured by present-day experiment, and our primary tool in this effort will be {\tt{smelli}} \cite{Aebischer:2018iyb}, an open-source python package which builds upon the  {\tt{flavio}} \cite{Straub:2018kue} and {\tt{Wilson}} \cite{Aebischer:2018bkb} programs. The goal of {\tt{flavio}} is to enable the automated calculation of flavoured processes  in terms of dimension-six Wilson coefficients $C$ of the SM effective field theory (SMEFT, valid above the EW scale) or the WET (valid below the EW scale). This package includes a large library of experimental measurements in the flavour and EW sectors. In addition, {\tt{Wilson}} computes the renormalization group evolution (RGE)  of SMEFT operators, their matching onto the WET at relevant scales, and the further RGE running of WET operators to hadronic scales that are (often) of interest in flavour physics. Augmenting these capabilities, {\tt{smelli}} computes a global likelihood function in the space of SMEFT coefficients, i.e. the object   
\begin{equation}
\label{eq:smelliLIKE}
L_\text{SMEFT}(\vec{C}) = \prod_{i} L_\text{exp} \left(\vec{O}^\text{exp}_i, \vec{O}^\text{th}_i \left(\vec{C}, \vec{\theta}\right) \right) \times L_\theta \left(\vec{\theta} \right) \,.
\end{equation}
Here $L_\text{exp}$ are likelihood distribution functions which depend on independent experimental measurements $\vec{O}^\text{exp}_i$ with associated theory predictions $\vec{O}^\text{th}_i$, which are themselves functions of the Wilson coefficients $\vec{C}$ and additional model-independent phenomenological parameters $\theta$ (e.g. form factors and decay constants). Then $L_\theta$ accounts for any experimental or theoretical constraints on these nuisance parameters. Note however that the actual {\tt{smelli}} implementation of \eqref{eq:smelliLIKE} relies on a `nuisance-free' approximation to the total global likelihood function, which effectively `integrates out' the theoretical errors associated to $\theta$, treating them as additional experimental uncertainties.  This is achieved by a factorization of $L_\text{SMEFT}$ into likelihoods for observables that a) have negligible theoretical vs. experimental uncertainty or b) that have reliably Gaussian theory and experimental uncertainties, and where the former only weakly depends on $\vec{C}$ and $\vec{\theta}$. While care must be taken for certain observables (e.g. CKM angles) that do not necessarily respect these assumptions, they are reliable and frequently employed (see e.g. \cite{Efrati:2015eaa,Falkowski:2017pss,Altmannshofer:2014rta,Descotes-Genon:2015uva}) for the analyses we are attempting here.
Finally, {\tt{smelli}} also includes the one-loop RGE that mix flavour structures in the SMEFT, and which are required for a consistent matching to the WET, where QCD and QED renormalization is flavour-blind \cite{Jenkins:2013zja,Jenkins:2013wua,Alonso:2013hga,Aebischer:2015fzz,Jenkins:2017jig,Aebischer:2017gaw,Jenkins:2017dyc}. This matching and associated RGE is performed automatically in {\tt{smelli}}, allowing coherent comparisons to low-energy experimental data given a UV new physics scale $\Lambda$.  For a more complete description of {\tt{smelli}} functionality, its built-in assumptions, and an exhaustive list of observables included in its likelihoods, see \cite{Aebischer:2018iyb} and the documentation at \url{https://github.com/smelli/smelli}.

%%%%%%%%%%%%%%%%%%%%%%%%%%%%%%%%%%%%
\subsection{Matching the Vector Leptoquark Singlet}
In order to perform a ${\tt{smelli}}$ analysis for our SMFL, we first recall the tree-level SMEFT matching onto the LL operators of the $\Delta^\mu$ vector LQ singlet given in \eqref{eq:LLyukSU2}. When computed at the new physics scale $\Lambda = M_\Delta$ it yields (see e.g. \cite{deBlas:2017xtg}):
\begin{equation}
\label{eq:treeC}
\left[C_{LQ}^{(1)}\right]_{ijkl} = \left[C_{LQ}^{(3)}\right]_{ijkl} = -\frac{\lambda_{LQ}^{kj} \, \lambda_{LQ}^{li\star}}{2 M_\Delta^2} \,,   
\end{equation}
which are the Wilson coefficients of the dimension-six four-fermion SMEFT operators given by
\begin{equation}
\label{eq:fourfermions}
\left[\mathcal{O}^{(1)}_{LQ}\right]_{ijkl} = \left(\overline{L}_i \gamma_\mu L_j \right)\left(\overline{Q}_k \gamma_\mu Q_l \right), \,\,\,\,\,\,\,\,\,\, \left[\mathcal{O}^{(3)}_{LQ}\right]_{ijkl} = \left(\overline{L}_i \gamma_\mu \tau^I L_j \right)\left(\overline{Q}_k \gamma_\mu \tau^I Q_l \right).
\end{equation}
In addition to this tree-level matching, we follow \cite{Aebischer:2019mlg} and include additional one-loop matching contributions to quark dipole operators in the SMEFT, which can generate the known \cite{Crivellin:2018yvo} vector LQ singlet matching contributions to electric and chromomagnetic dipole operators in the low-energy WET.  The relevant dimension-six SMEFT operators are 
\begin{equation}
\nonumber
\left[\mathcal{O}_{dB} \right]_{ij} = \left(\overline{Q}_i \sigma^{\mu \nu} d_j \right) \phi\, B_{\mu \nu} \,, \,\,\,\,\,\, \left[\mathcal{O}_{dW} \right]_{ij} = \left(\overline{Q}_i \sigma^{\mu \nu} d_j \right) \tau^I \phi\, W^I_{\mu \nu}\,, \,\,\,\,\,\, \left[\mathcal{O}_{dG} \right]_{ij} = \left(\overline{Q}_i \sigma^{\mu \nu} T^A d_j \right) \phi \,G^A_{\mu \nu}\,,
\end{equation}
which are catalogued alongside all remaining independent dimension-six operators in \cite{Grzadkowski:2010es}.  Note that in {\tt{smelli}} the Warsaw basis of \cite{Grzadkowski:2010es} (the {\tt{flavio}} basis of \cite{Straub:2018kue}) is the default basis when obtaining likelihoods in the SMEFT (WET).
Again computing the relevant matching at $\Lambda = M_\Delta$, one finds \cite{Aebischer:2019mlg}
\begin{align}
\nonumber
\left[C_{dW}\right]_{23} &= \frac{Y_b}{6} \frac{g}{16 \pi^2} \frac{\lambda_{LQ}^{2i} \lambda_{LQ}^{3i\star}}{M_\Delta^2} \, , \,\,\,\,\,\,\,\,\left[C_{dB}\right]_{23} = -\frac{4Y_b}{9} \frac{g^\prime}{16 \pi^2} \frac{\lambda_{LQ}^{2i} \lambda_{LQ}^{3i\star}}{M_\Delta^2}\, , \,\,\,\,\,\,\,\,\left[C_{dG}\right]_{23} = -\frac{5 Y_b}{12} \frac{g_s}{16 \pi^2} \frac{\lambda_{LQ}^{2i} \lambda_{LQ}^{3i\star}}{M_\Delta^2}\, \\
\nonumber
\left[C_{dW}\right]_{32} &= \frac{Y_s}{6} \frac{g}{16 \pi^2} \frac{\lambda_{LQ}^{3i} \lambda_{LQ}^{2i\star}}{M_\Delta^2} \, , \,\,\,\,\,\,\,\,\left[C_{dB}\right]_{32} = -\frac{4Y_s}{9} \frac{g^\prime}{16 \pi^2} \frac{\lambda_{LQ}^{3i} \lambda_{LQ}^{2i\star}}{M_\Delta^2}\, , \,\,\,\,\,\,\,\,\left[C_{dG}\right]_{32} = -\frac{5 Y_s}{12} \frac{g_s}{16 \pi^2} \frac{\lambda_{LQ}^{3i} \lambda_{LQ}^{2i\star}}{M_\Delta^2} \,. 
\\
\label{eq:oneloopC}
\end{align}
Here the lepton index $i$ is summed over.  In addition to the Lagrangian conventions discussed above Section \ref{sec:SMFL}, these were also computed in the limit of a diagonal down-quark Yukawa matrix, with $Y_{s,b}$ the respective Yukawa couplings to strange and bottom quarks.  Note that, as mentioned above, one expects logarithmic divergences to appear in dipole processes in the `minimal' coupling scenario where $k = 1$ in \eqref{eq:DeltaLagrangian} \cite{Ferrara:1992yc,Barbieri:2015yvd}, and therefore the operators in \eqref{eq:oneloopC} may not be induced in a sensible UV matching with this parameter choice. It is also clear that, when $k = 1$, triple-vector couplings between $\Delta-\Delta-B/G$ are not present at tree level, and therefore the one-loop diagram leading to non-zero $C_{dG}$ (e.g.) is not present.  Regardless, we note that our analysis in Section 4 is largely insensitive to these UV details, as we have found that turning off all of the dipole operators in \eqref{eq:oneloopC} only results in a roughly 1$\%$ correction to the best-fit ratio of $\lambda_{s\tau}/\lambda_{b\tau}$ when $M_\Delta = 1$ TeV (e.g.).  In what follows we therefore show fit results in the `full' $k = 0$ scenario.

Finally, we recall that as with other similar studies, we have chosen to set the RR couplings $x^{RR}$ and $x^{\overline{RR}}$ in \eqref{eq:LLyukSU2} to zero.  As can be deduced from model-independent EFT fits (see e.\,g.\ \cite{Murgui:2019czp,Shi:2019gxi,Aebischer:2019mlg,Blanke:2018yud,Blanke:2019qrx,Iguro:2020cpg}) and as will be seen explicitly below, these RH couplings are not necessary in minimal explanations of the observed LFU anomalies, and it has been further shown \cite{Baker:2019sli} that exclusion limits on $M_\Delta$ strengthen when $x^{RR} \neq 0$.  Hence \eqref{eq:treeC}-\eqref{eq:oneloopC} represent the complete set of relevant SMEFT Wilson coefficients implemented in our {\tt{smelli}} analysis.

%%%%%%%%%%%%%%%%%%%%%%%%%%%%%%%%%%%%
\subsection{Scanning the Vector Leptoquark Singlet}
\label{sec:SCANS}

Given \eqref{eq:treeC}-\eqref{eq:oneloopC}, one is in a position to scan over the SMFL couplings $\lambda_{dl}$ and allow {\tt{smelli}} to compute likelihoods at each phase-space point.  One can collect this information as a function of  $\lambda_{dl}$ and determine the experimentally favoured space of couplings for a given SMFL pattern/model, as well as the observables which contribute the most significant pulls.  
\\
\\
\\
Specifically, in performing our scans we
\begin{enumerate}
    \item take the LQ couplings $\lambda_{dl}$ to be real.  As will be seen, there is ample parameter space of interest even without additional complex degrees of freedom.
    \item build an array of $\lbrace \lambda_{sl}, \lambda_{bl} \rbrace$ by dividing the phase space in either dimension by a predefined set of intervals.  We then calculate the Wilson coefficients in \eqref{eq:treeC}-\eqref{eq:oneloopC} for all points on this grid.
    \item perform a global likelihood analysis using {\tt{smelli v2.3.2}}\footnote{In a prior version of this paper we utilized {\tt{smelli v2.0.0}}, which was released in December 2019 and therefore did not include a number of code improvements, nor a host of recent experimental results, including (e.g.) the 2021 measurement of $\mathcal{R}_K$ \cite{LHCb:2021trn}.  We thank an anonymous referee for pointing this out to us.} at each parameter point on the $\lambda_{dl}$ grid.  This is performed by calling {\tt{smelli.GlobalLikelihood()}}, which we also modify using the {\tt{custom\_likelihoods}} attribute.  This latter functionality allows us to define custom sets of observables contributing to a likelihood computation.
    \item collect all likelihoods computed and determine the values associated to the minimum $\Delta\chi^2$ for a given pattern (and a given set of observables).  We do so by calling the {\tt{log\_likelihood\_global()}} method, which returns $\Delta \log L = -\Delta \chi^2 /2$, where $\Delta \chi^2$ is the BSM $\chi^2$ minus its SM value.  We use this to then compute 1$\sigma$ and 2$\sigma$ likelihood contours about the $\Delta \chi^2$ minimum.
    \item in order to determine the relative pull of any given set of observables, we also use the {\tt{log\_likelihood\_dict()}} method, which returns the dictionary of all contributions to $ \Delta\log L$ from the individual products in (the {\tt{smelli}} implementation of) \eqref{eq:smelliLIKE}.  Note that in addition to the classes of observables already segregated in {\tt{smelli}} through the inclusion of separate internal {\tt{YAML}} files, any of the {\tt{custom\_likelihoods}} we defined ourselves will also be given as independent contributions.    
\end{enumerate}
We now report the results of these scans for SMFL patterns of distinct phenomenological interest:  the lepton isolation patterns 
$\lambda_{dl}^{[e,\mu,\tau]}$. 
We also report the relative pull of {\bf{$\mathcal{R}_{K^{(*)},D^{(*)}}$}} (at the global $\Delta\log L$ maximum), which are especially interesting due to their present deviations from SM predictions.

\begin{table}
\centering
{\renewcommand{\arraystretch}{1.4}
\begin{tabular}{|c||c|c|c|c|c|}
\hline
SMFL & $M_\Delta$ & Best Fit $\left(\lambda_{sl}, \lambda_{bl}\right)$ & $ \Delta \log L \vert_{\mathcal{R}_{D^{(\star)}}}$ &
$ \Delta \log L \vert_{\mathcal{R}_{K^{(\star)}}}$ &
 $ \Delta \log L \vert_\text{Global}$  \\
\hline
\hline
\multirow{2}{*}{$\lambda_{dl}^{[\tau]}$} & 2 TeV  & $\left( 0.64, 0.72\right)$ & 12.638 &
\multirow{2}{*}{\text{N.A.}} &
18.457 \\
& 1 TeV & $\left( 0.30, 0.38\right)$ & 12.622  & &  17.970
\\
\hline
\hline
\multirow{2}{*}{$\lambda_{dl}^{[\mu]}$}  & 2 TeV & $\left( 0.147, -0.021\right)$ & 0.107 & 6.535 & 
20.648 \\
& 1 TeV & $\left( 0.078, -0.01\right)$ & 0.107  & 6.469 &  20.812 \\
\hline
\hline
\multirow{2}{*}{$\lambda_{dl}^{[e]}$} & 2 TeV & $\left( 0.016, 0.172\right)$ & -$\mathcal{O}(10^{-2})$ & 7.306 & 
8.222 \\
& 1 TeV & $\left( 0.006, 0.114\right)$ & -$\mathcal{O}(10^{-2})$  & 7.302 &  8.224 \\
\hline
\end{tabular}}
\caption{
Results from the $M_\Delta = \lbrace 1,2 \rbrace $ TeV two-parameter likelihood ($ \Delta \log L = - \Delta \chi^2 /2$) scans of Section \ref{sec:SCANS}, including the individual contributions of the (potentially) anomalous observables $\mathcal{R}_{K^{(\star)},D^{(\star)}}$.  Column 3 gives the best-fit values of $\left(\lambda_{sl}, \lambda_{bl}\right)$, corresponding to the global likelihood maximum of the scans, found in column 6, which consider all available data in {\tt{smelli}}.  Columns 4-5 then give the individual contributions of $\mathcal{R}_{K^{(\star)},D^{(\star)}}$ to this likelihood (again at the best-fit coordinates).  See the text and Figure \ref{fig:SmelliScans} for more details.
\label{tab:largeL}
}
\end{table}
%-------------------------------------------------------------------------

%------------------------------------------------------------------------
\begin{figure}[tp]
\includegraphics[width = 0.5 \textwidth]{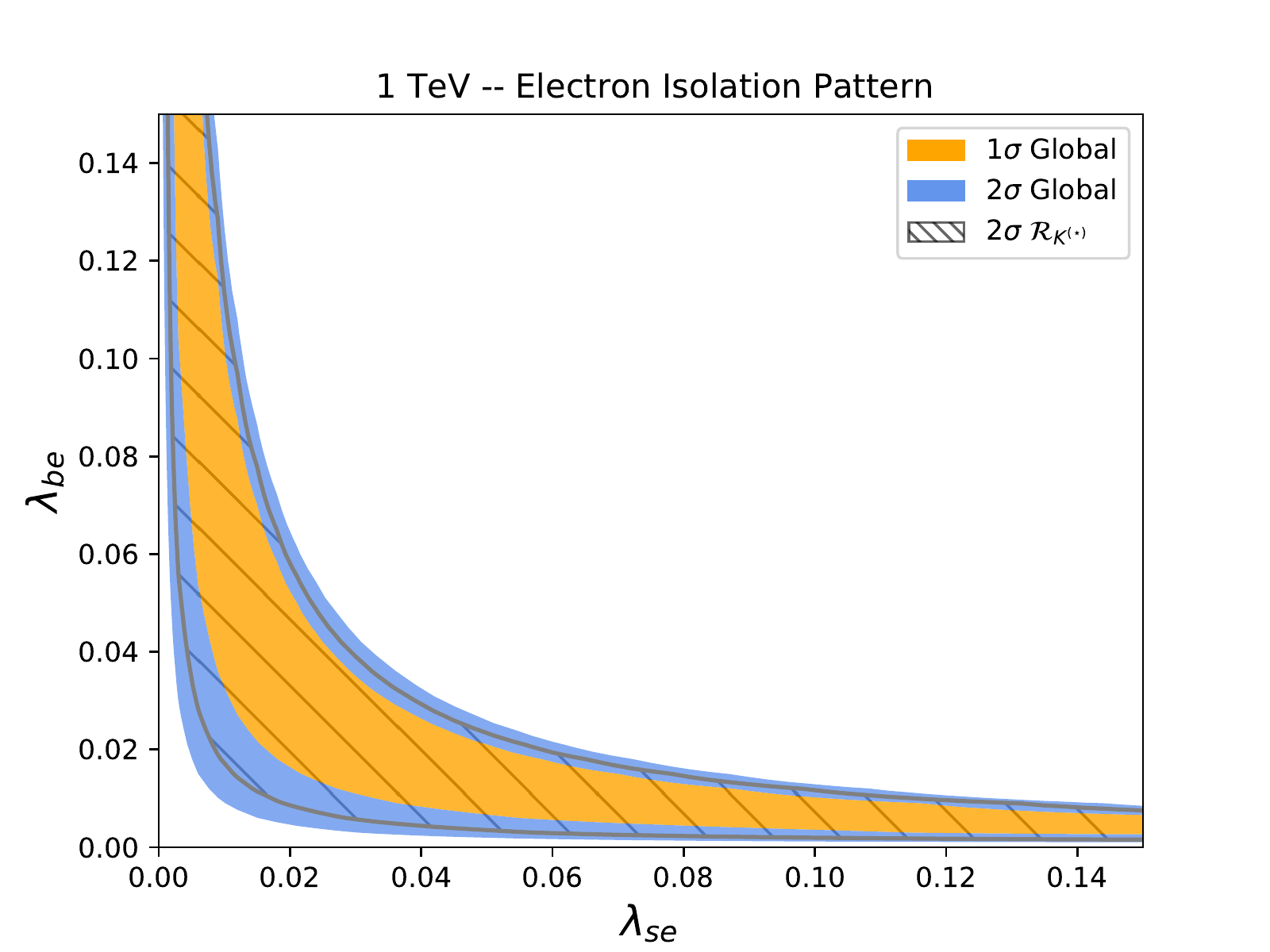}
  \includegraphics[width = 0.5 \textwidth]{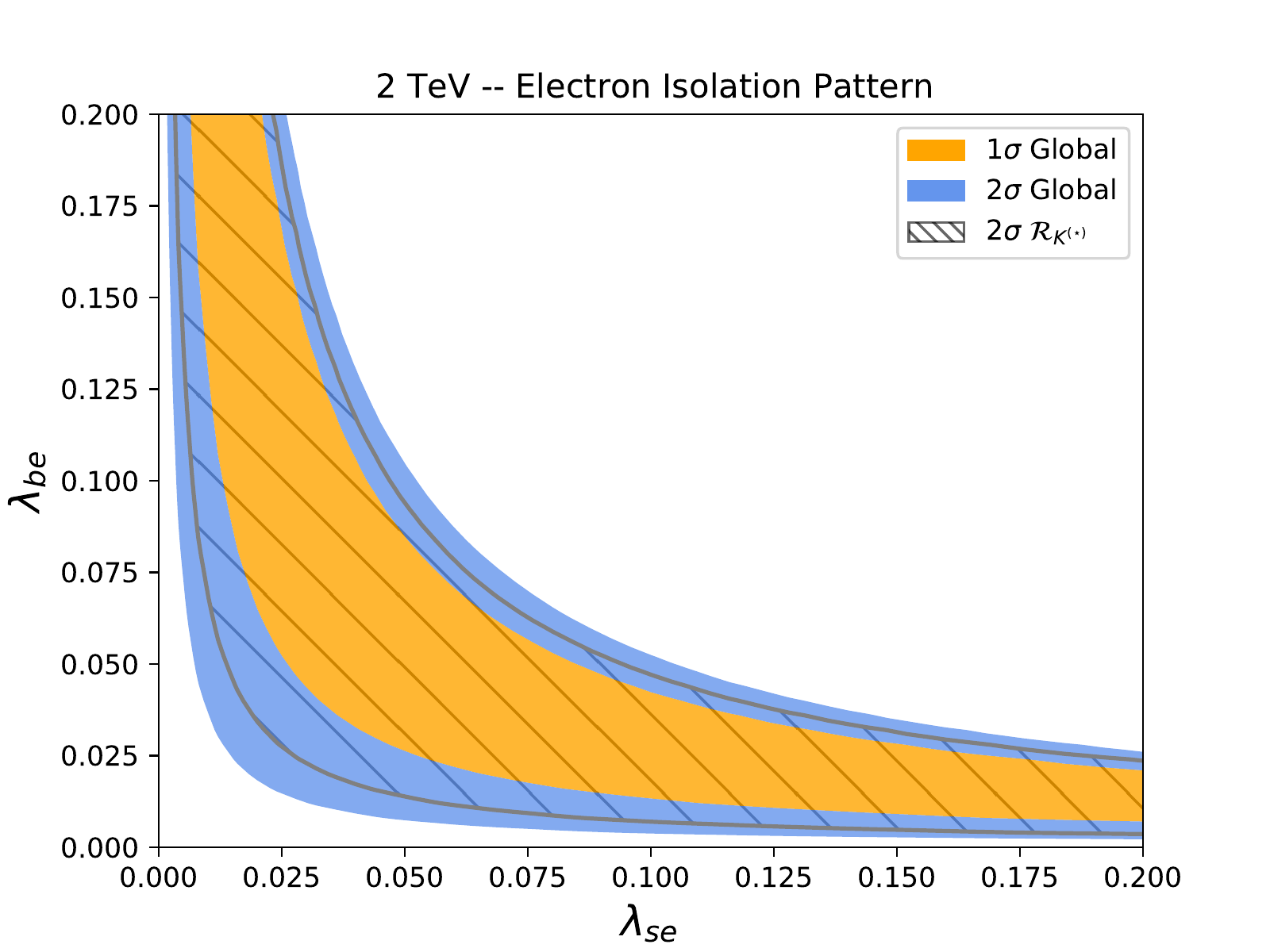} \\
  \includegraphics[width = 0.5 \textwidth]{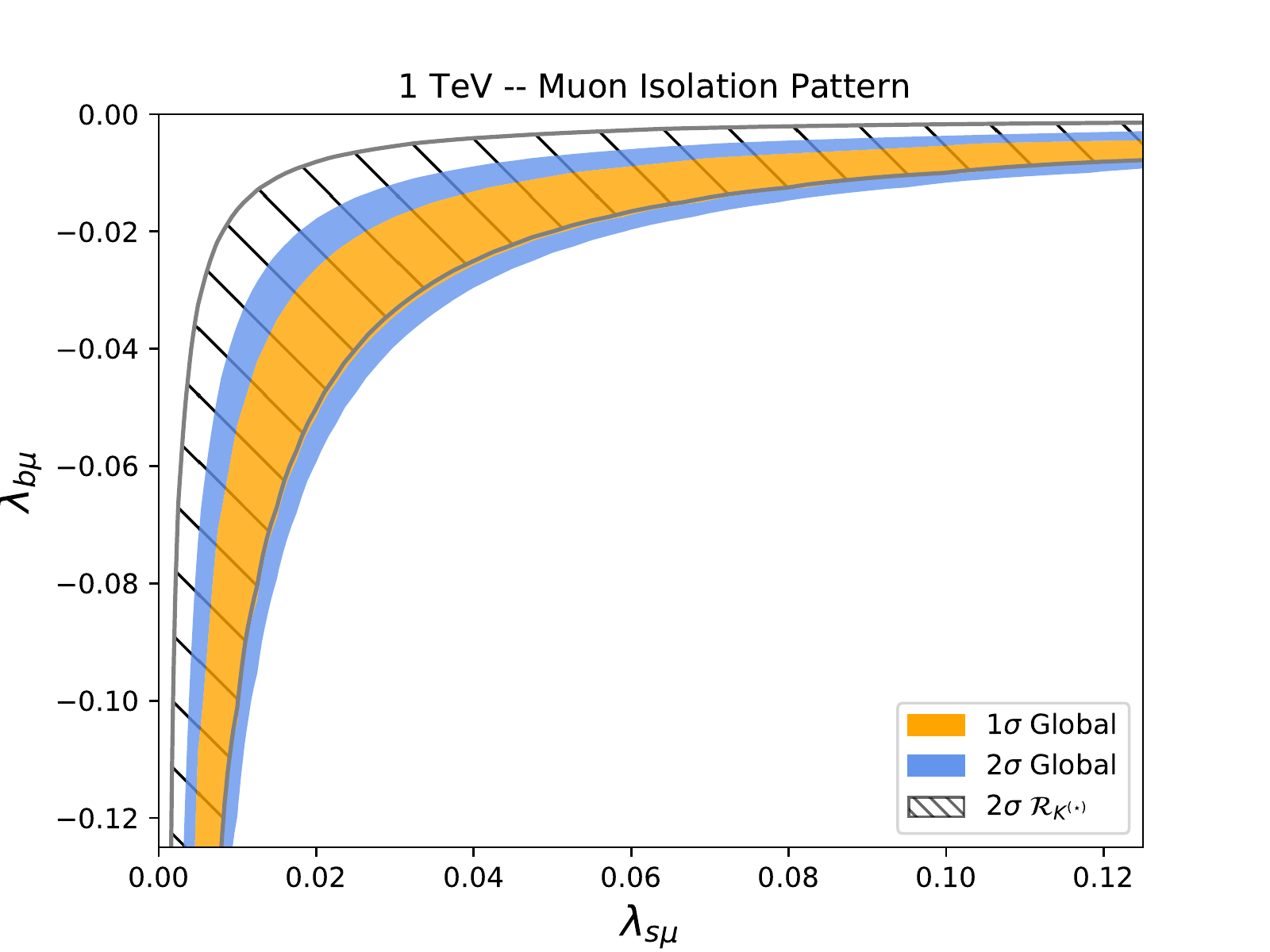}
  \includegraphics[width = 0.5 \textwidth]{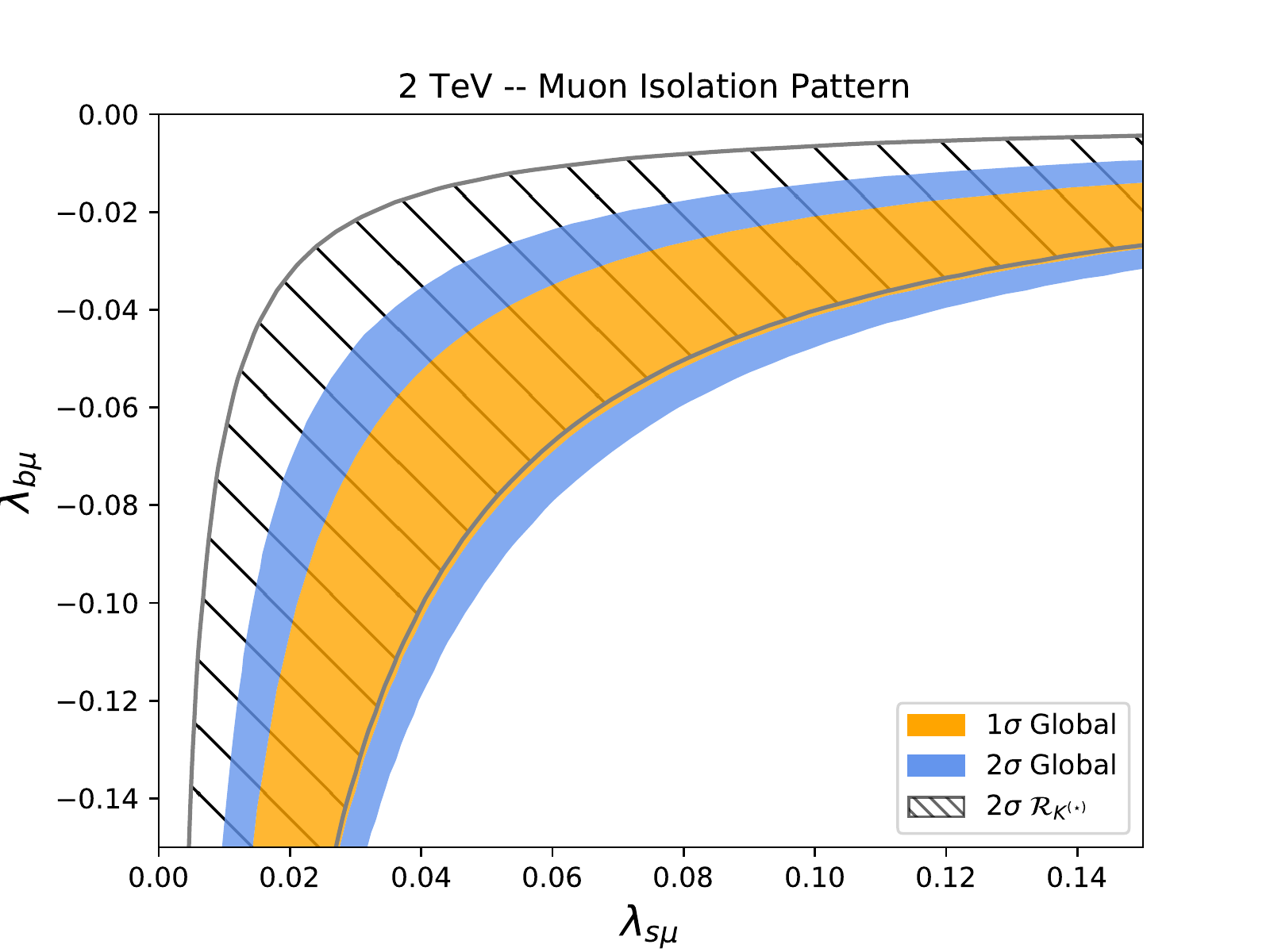}\\
    \includegraphics[width = 0.5 \textwidth]{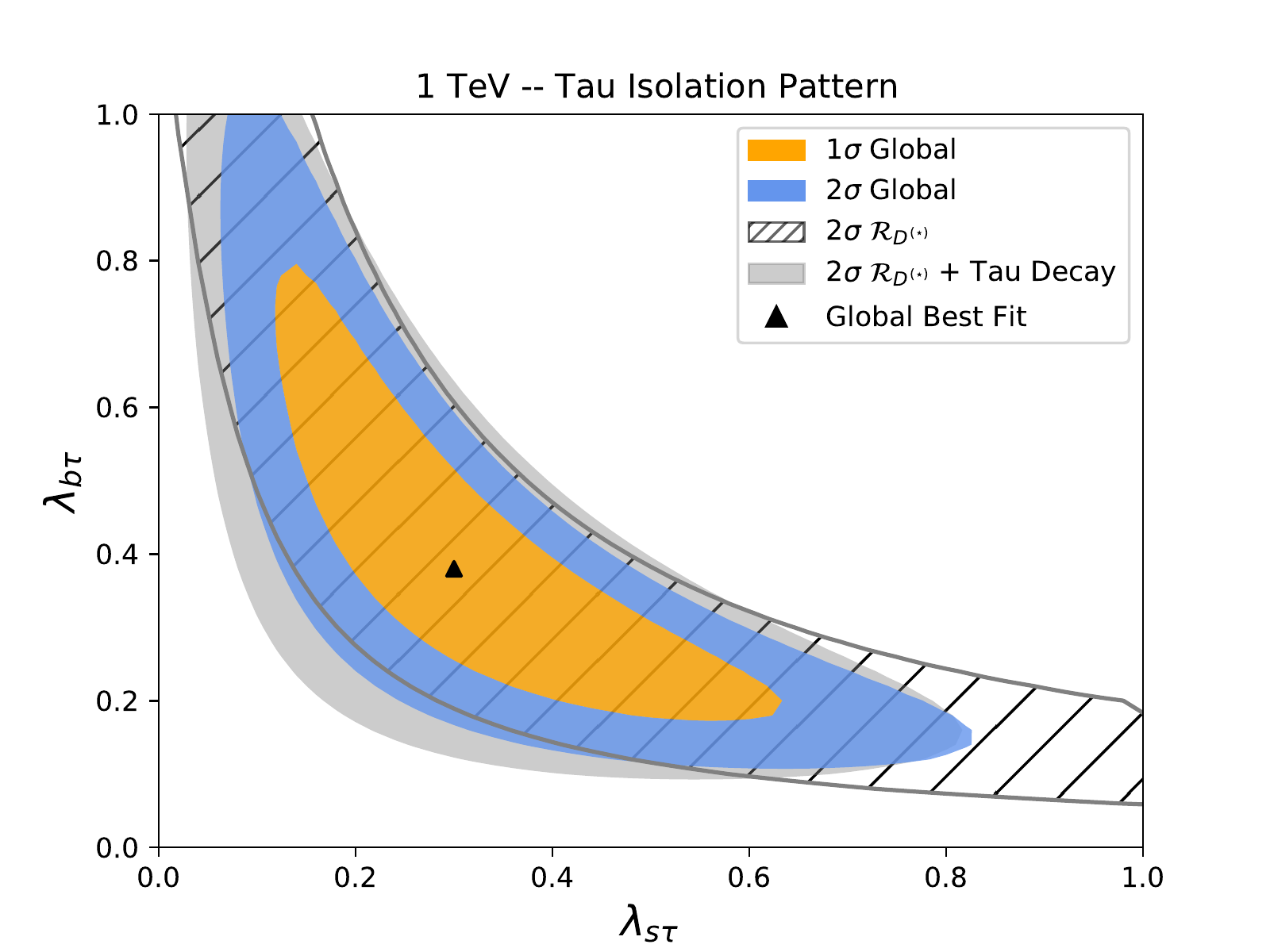}
  \includegraphics[width = 0.5 \textwidth]{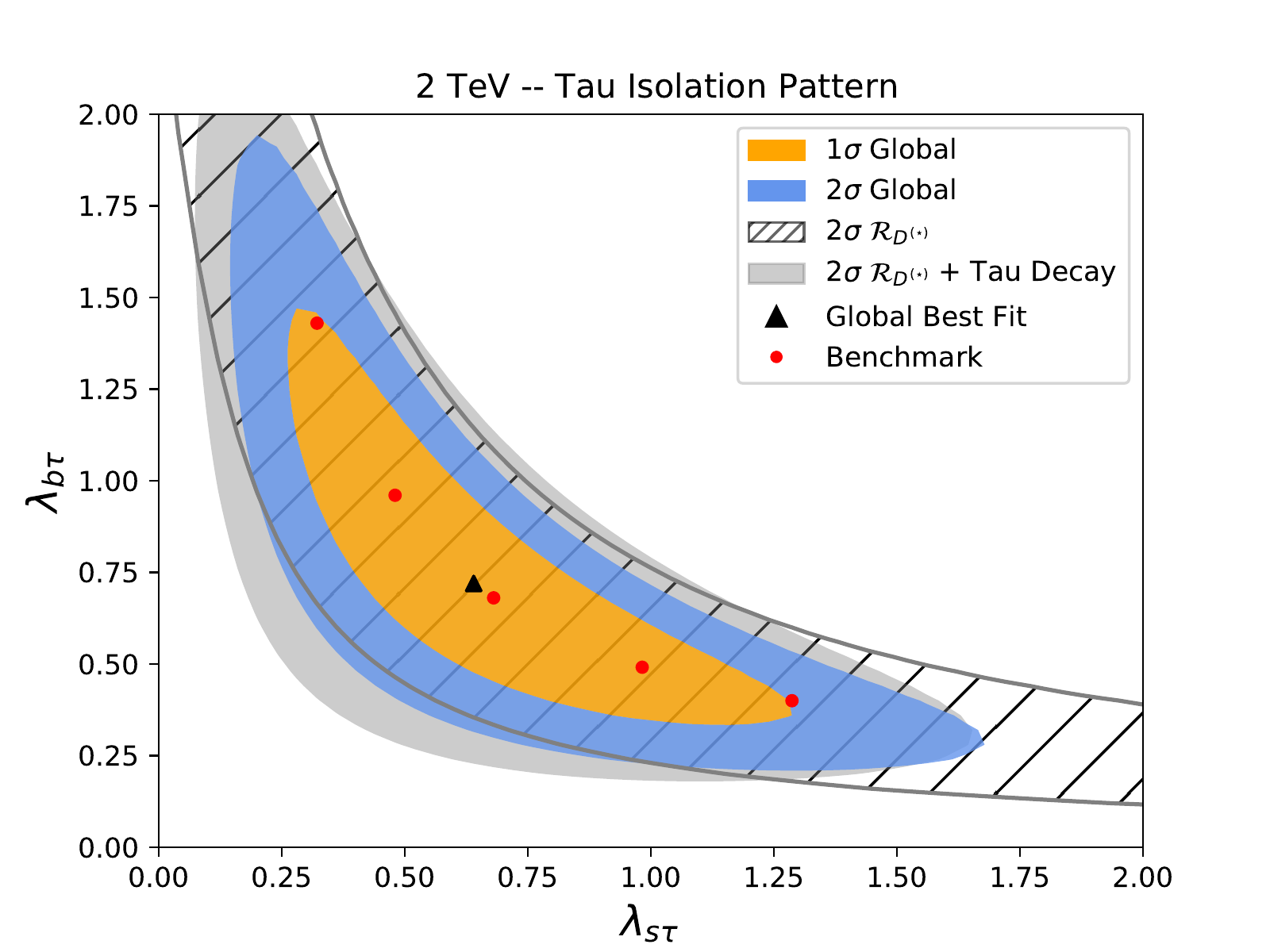}
  \caption{~\label{fig:SmelliScans} {\bf{Top Left:}} Likelihood contours for the two-parameter $e$-isolation pattern $\lambda_{dl}^{[e]}$ at $M_\Delta = 1$ TeV, including  $2\sigma$ contours from  $\mathcal{R}_{K^{(\star)}}$ constraints alone (hatched region), as well as global $1\sigma$ (orange region) and $2\sigma$ (blue region) preferred contours considering all data available to {\tt{smelli}}. {\bf{Top Right:}}  The same, but for $M_\Delta = 2$ TeV.  {\bf{Middle Row:}}  The same, but for the $\mu$-isolation pattern $\lambda_{dl}^{[\mu]}$.  {\bf{Bottom Row:}}  The same, but for the $\tau$-isolation pattern $\lambda_{dl}^{[\tau]}$, and including $2\sigma$ contours from  $\mathcal{R}_{D^{(\star)}}$ constraints alone (hatched region), and $\mathcal{R}_{D^{(\star)}} + \tau$ decay modes (gray region). Here the global best-fit (black, triangular) and benchmark (red, circular) points in $\lbrace \lambda_{s\tau}, \lambda_{b\tau} \rbrace$ we obtained from our scan are also shown.
 }
\end{figure}
%------------------------------------------------------------------------

%%%%%%%%%%%%%%%%%%%%%%%%%%%%%%%%%%%%%%%%%%%%%%%%%%%%%%%%%%%%%%%%%%%%%%%
\subsubsection*{Electron and Muon Isolation Patterns}

We first investigate $\lambda_{dl}^{[e,\mu]}$ from \eqref{eq:yukeisolation}, the e- and $\mu$-isolation patterns, where the results of the {\tt{smelli}} scans we performed as described above are given in the top (middle) panels of Figure \ref{fig:SmelliScans} for the electron (muon) patterns, as well as numerically in Table~\ref{tab:largeL}.  One observes in Figure~\ref{fig:SmelliScans} that a broad range of couplings is allowed at the 1$\sigma$ and 2$\sigma$ confidence level, considering all available data, and that the potentially anomalous measurement of $\mathcal{R}_{K^{(\star)}}$ is well-described in these models.  Indeed, the overall shape of the global parameter space preferred largely follows that preferred by $\mathcal{R}_{K^{(\star)}}$ alone. Note that the 2021 measurement of $\mathcal{R}_K$ \cite{LHCb:2021trn} increased the tension with the SM, implying a stronger preference for non-zero LQ couplings $\lambda_{qe}$ and $\lambda_{q\mu}$ (with $q=b,s$). 
The additional pull to non-zero couplings in the $\mu$-isolation pattern originates in the so-called $b\to s\mu^+\mu^-$ anomalies observed in the decays $B\to K^*\mu^+\mu^-$, $B_s\to\phi\mu^+\mu^-$ etc. that can not be addressed in the electron-isolation pattern. 
 In the latter case, non-zero couplings have received further support from the recent update of the $B^0\to K^{*0}\mu^+\mu^-$ angular analysis \cite{LHCb:2020lmf}, the newly measured $B^+\to K^{*+}\mu^+\mu^-$ angular observables \cite{LHCb:2020gog}, and the recent experimental update of the  $B_s\to\phi\mu^+\mu^-$ branching ratio \cite{LHCb:2021zwz}. We refer the reader to \cite{Altmannshofer:2021qrr} for further details on the  implications of these measurements. 
Additionally, as seen in Table \ref{tab:largeL}, the global best-fit value is also favoured over SM couplings alone as it slightly softens  the charged-current LFU anomaly $\mathcal{R}_{D^{(\star)}}$, at least for the $\mu-$isolation pattern.  When considering only $\mathcal{R}_{D^{(\star)}}$ data, better likelihoods can be obtained in the parameter space scanned.  $\lambda_{dl}^{[\mu]}$'s ability to resolve LFU anomalies whilst generating a distinct collider phenomenology has been known for some time (see e.g. \cite{Hiller:2018wbv,Hiller:2021pul}).

As a final note, we have observed that the overall $-\Delta \chi^2/2$ likelihood distributions for both $\lambda_{dl}^{[e,\mu]}$ are somewhat flat, in that multiple points spanning a broad domain in the two-dimensional contours presented fall very close to the likelihood found for the global best-fit point.  For example, we can identify a number of parameter space points whose likelihoods are within a percent (or less) of the global maximum, but whose coordinates are a factor of 2 (or more) away from those presented in Table \ref{tab:largeL}. We found this behavior for  $\lambda_{dl}^{[e,\mu]}$ at both $M_\Delta = \lbrace 1, 2 \rbrace$ TeV, and it is for this reason that we do not find it instructive to plot a `global' best-fit point in Figure \ref{fig:SmelliScans}, but instead give this information in Table \ref{tab:largeL}, to illustrate the overall quality of the fits. 

%%%%%%%%%%%%%%%%%%%%%%%%%%%%%%%%%%%%%%%%%%%%%%%%%%%%%%%%%%%%%%%%%%%%%%%
\subsubsection*{Tau Isolation Pattern}

We next investigate $\lambda_{dl}^{[\tau]}$ from \eqref{eq:yukeisolation}, the $\tau$-isolation pattern.  This texture has been studied before in the context of a toy vector singlet LQ model \cite{Aebischer:2019mlg} (although without the SMFL symmetry-based motivation for its flavour structure), and here we update those results given improved experimental and theoretical developments over the last years.\footnote{We thank Peter Stangl for pointing out (e.g.) \cite{Bordone:2019vic}, whose updated $B \rightarrow D^{(\star)}$ form factors impact predictions (and uncertainty estimates) for $\mathcal{R}_{D^{(\star)}}$. Note also that the gray-shaded contour of the bottom-right panel of Figure \ref{fig:SmelliScans} can be compared to the corresponding contours in Figure 6 of \cite{Aebischer:2019mlg}, where we find good qualitative agreement, given the updates in the code and both experimental and theoretical inputs.}  

Using the algorithm described in Section \ref{sec:SMELLI}, we produce the bottom panels in Figure \ref{fig:SmelliScans}.  The graphics illustrate the contours contributing to the global likelihood coming from the ratio observables $\mathcal{R}_{D^{(\star)}}$ (hatched region), those from a custom fit combining both $\mathcal{R}_{D^{(\star)}}$ and the leptonic $\tau$-decay modes $\text{BR}\left(\tau^- \rightarrow l^- \nu \bar{\nu}\right)$ with $l = e, \mu$, $\text{BR}\left(\tau^+ \rightarrow K^+ \bar{\nu}\right)$, and $\text{BR}\left(\tau^+ \rightarrow \pi^+ \bar{\nu}\right)$ (gray region), and finally the global $1\sigma$ (orange region) and $2\sigma$ (blue region) preferred contours upon considering all experimental datasets in {\tt{smelli}}. Note that leptonic $\tau$ decays and $\mathcal{R}_{D^{(\star)}}$ were identified as dominant contributors to the overall likelihoods for  $\lambda_{dl}^{[\tau]}$ in \cite{Aebischer:2019mlg}, and we confirm that observation in our analysis. 
 We give these at $M_\Delta = \lbrace 1, 2\rbrace$ TeV (left and right panels, respectively), and the best-fit $\left(\lambda_{s\tau},\lambda_{b\tau}\right)$ values are shown in black, where the global likelihood maxima  of our scans are realized.  The numerical values of these coordinates as well as the maximum $\Delta \log L$ is again given in Table \ref{tab:largeL}.  
 For the 2 TeV contours we also plot `benchmark points' in red that we will use in the upcoming collider analysis of Section \ref{sec:LHC}.  
 
Finally, we observe from Table \ref{tab:largeL} that the overall global likelihood for $\lambda_{dl}^{[\tau]}$ is significantly larger than that found when considering $\mathcal{R}_{D^{(\star)}}$ alone.  This effect has been observed before, and can be traced back to gauge-induced one-loop renormalization group mixing that generates a lepton-universal contribution to the semi-leptonic operators of \eqref{eq:fourfermions} (above and below the EW scale), from the non-universal, $\tau$-specific semi-leptonic operator matched at tree-level (cf. \eqref{eq:treeC} for the $\lambda_{dl}^{[\tau]}$ SMFL)--- see e.g. the discussion in \cite{Aebischer:2019mlg,Crivellin:2018yvo}.  The operator mixing therefore makes this $\tau$-isolation pattern sensitive to other semileptonic processes, e.g. $B \longrightarrow K^{(\star)} \ell \ell$ branching ratios and angular observables, measurements of which are considered in the global likelihood.\footnote{We again thank an anonymous referee for this comment.}

\subsubsection*{Summary}

In conclusion, it is clear that the $\mu$- and the $\tau$-isolation patterns both provide excellent fits to the available data across a broad range of parameter space, and are able to solve the LFU anomalies $\mathcal{R}_{K^{(*)}}$ and $\mathcal{R}_{D^{(*)}}$, respectively. While the electron isolation pattern can also successfully address $\mathcal{R}_{K^{(*)}}$, it falls short of explaining the related anomalies in $b\to s \mu^+\mu^-$ transitions, and is hence less motivated from a phenomenological perspective.

In what follows we will further pursue the analysis of the $\tau$-isolation pattern and study its LHC signatures. A high-$p_T$ study of this scenario is particularly motivated, since we have seen that large LQ couplings are required by the global fit. In turn this precludes the possibility of evading direct searches by simply raising the LQ mass $M_\Delta$ beyond the reach of the LHC.

%%%%%%%%%%%%%%%%%%%%%%%%%%%%%%%%%%%%%%%%%%%%%%%%%%%%%%%
\section{Recasting LHC Exclusion Limits}
\label{sec:LHC}

\begin{figure}[!ht]
    \centering
    \includegraphics[scale = 0.5]{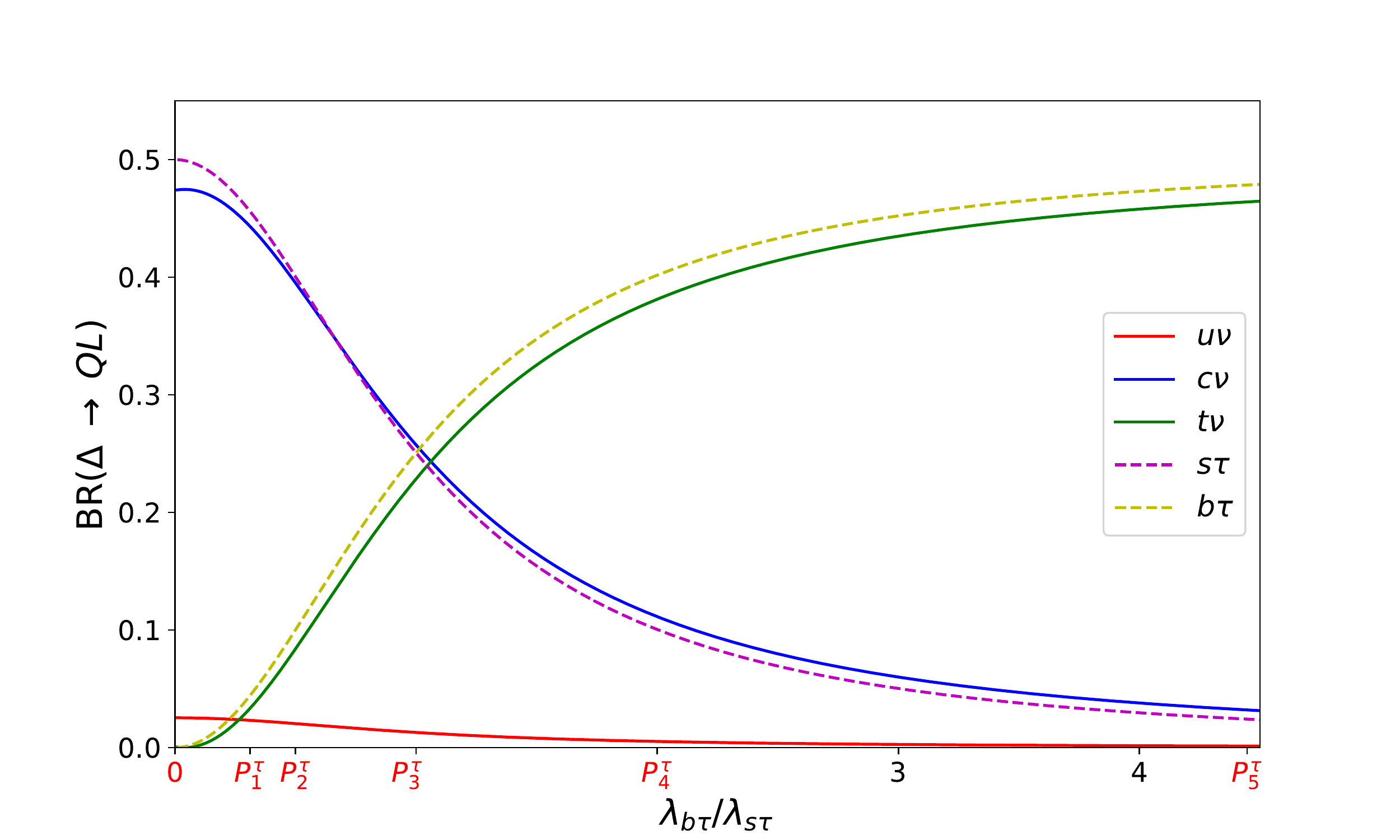} 
    \caption{2 TeV LQ branching ratios into quarks and leptons for the $\tau-$ isolation pattern. The channel magnitudes only depend on the ratio of the couplings. The non-vanishing $u-\nu$ contribution appears due to CKM rotations from $\lambda_{d\ell} \rightarrow \lambda_{u\nu}$. The other channels are (almost) exactly vanishing in the context of the $\tau-$isolation pattern. We have highlighted in red the position of the different benchmark points shown in Figure \ref{fig:SmelliScans}. Note that $P_{2}^\tau$, $P_{3}^\tau$ and $P_{4}^\tau$ have $\lambda_{b\tau}/\lambda_{s\tau}$ values of 0.5, 1 and 2, respectively.}
    \label{fig:BR_LQ_tauiso_2TEV}
\end{figure}

In this Section we consider existing direct searches published by ATLAS \cite{ATLAS:2021jyv} and reinterpret them in the context of our well motivated $\tau-$isolation scenario. In particular, both second- and third-generation quarks can be involved in the LQ decay. Assuming only left-handed couplings, and neglecting the fermion masses as well as CKM rotations, the different branching ratios exhibit the following structure
\newcommand{\BRsecond}{\ensuremath{\text{BR}_{\Delta}^{2^\text{nd} } } }
\newcommand{\BRthird}{\ensuremath{\text{BR}_{\Delta}^{3^\text{rd} } } }
\begin{gather}
        \BRthird = \text{BR}(\Delta \rightarrow b\tau) \simeq \text{BR}(\Delta \rightarrow t\nu), \\ 
        \BRsecond = \text{BR}(\Delta \rightarrow s\tau) \simeq \text{BR}(\Delta \rightarrow c\nu) \simeq 0.5 - \BRthird  \,, 
        \label{eq:BRstructure}
\end{gather}
where $\BRsecond$ (\BRthird) corresponds to the decay into the second (third) down-type quark. This structure holds very well as shown in Figure \ref{fig:BR_LQ_tauiso_2TEV} where the CKM and fermion mass effects have been taken into account with a 2 TeV mass for the LQ.  Using \eqref{eq:decaywidth} we obtain the following relation for the branching ratio as a function of the couplings: 
\begin{equation}
    \BRthird \simeq \frac{1}{2}\frac{\lambda_{b\tau}^2}{\lambda_{b\tau}^2 + \lambda_{s\tau}^2}.
    \label{eq:simplifiedBR}
\end{equation}

%------------------------------
\begin{table}[!t]
\centering
\begin{tabular}{cc}
{\renewcommand{\arraystretch}{1.5}
\begin{tabular}{|c|c|}
\hline
\multicolumn{2}{|c|}{$\lambda_{dl}^{[\tau]}$ Benchmark Points } \\
\hline
BP$_i$ & ($\lambda_{s\tau},\, \lambda_{b\tau}$) \\
      \hline
       BP$_1$ & (1.29, 0.4)  \\
       \hline
       BP$_2$ & ( 0.98,  0.49) \\
       \hline
       BP$_3$ & (0.68, 0.68)  \\
       \hline
       BP$_4$ & (0.48, 0.96) \\
       \hline
       BP$_5$ & (0.32, 1.43) \\
       \hline
\end{tabular}}
&
\includegraphics[width = .42\textwidth,valign=m]{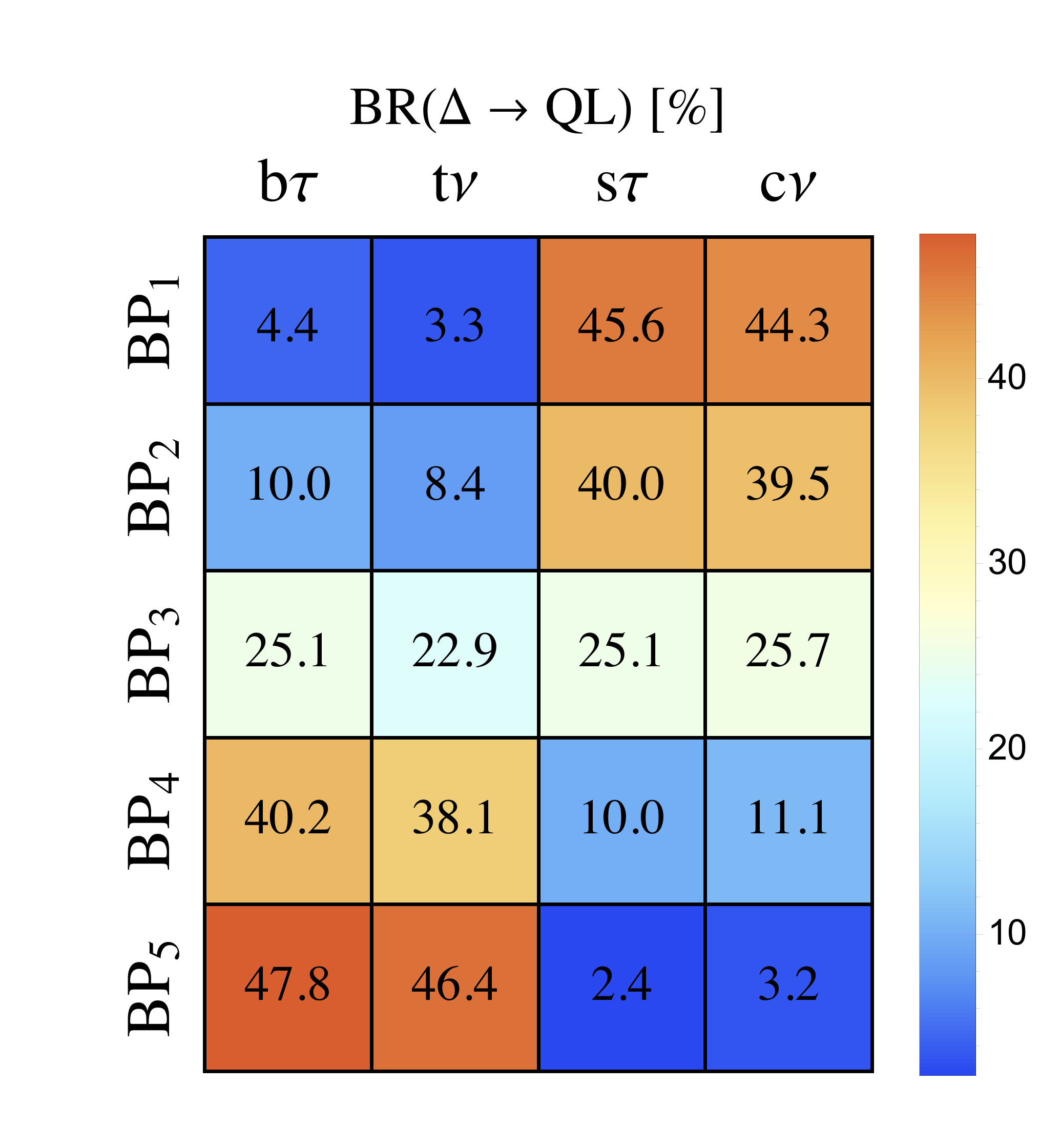}
\end{tabular}
\begin{tabular}{c}
\includegraphics[width = .88\textwidth]{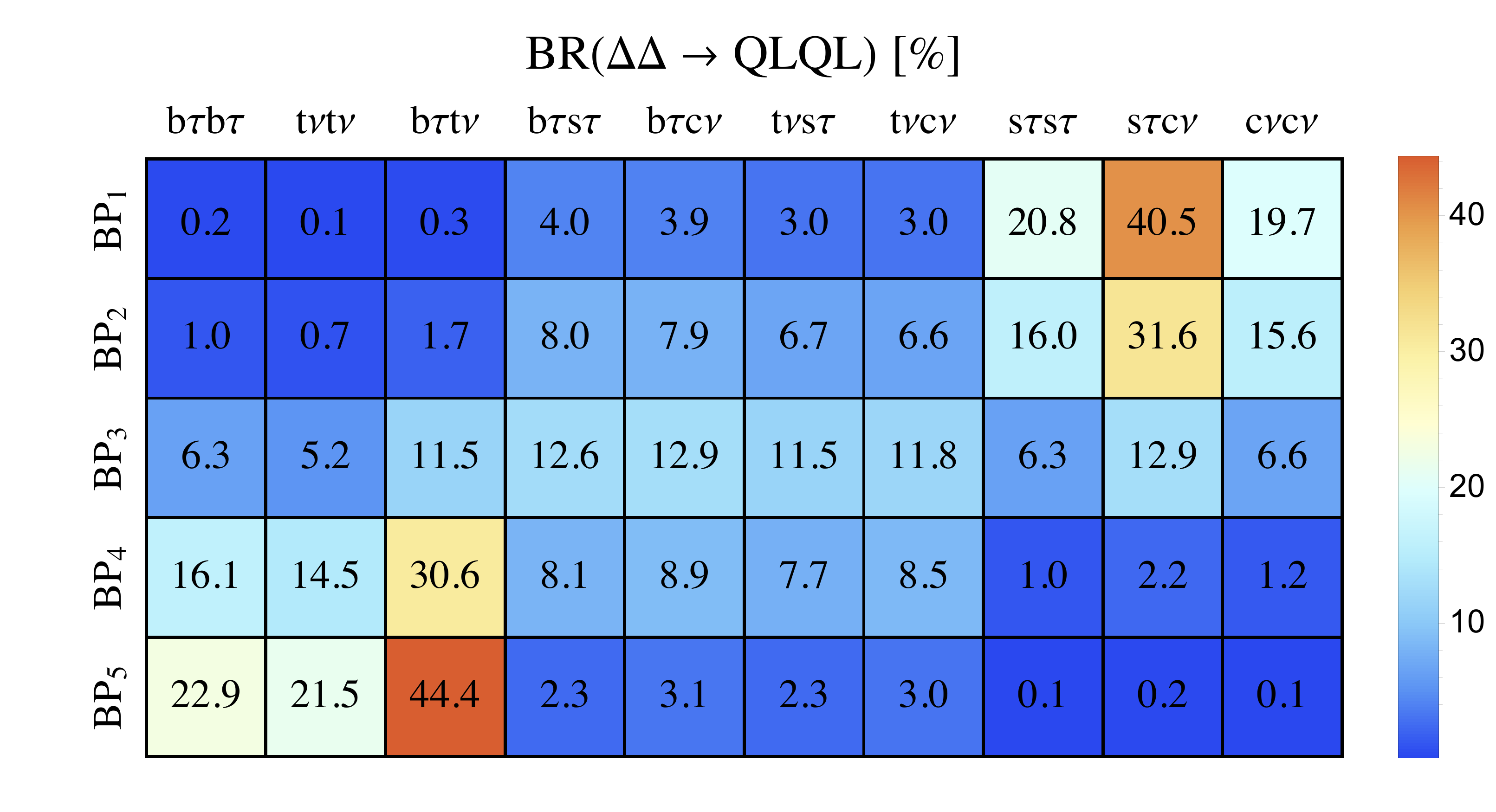}
\end{tabular}
  \caption{~\label{tab:BP_tau_iso} $\text{BR}(\Delta (\Delta) \rightarrow QL(QL))$ for 2 TeV LQ Benchmark Points from the $\lambda_{dl}^{[\tau]}$ scan of Section \ref{sec:SMELLI}. All asymmetric decay BR have an additional $\times 2$ factor because of permutation.}
\end{table}
%------------------------------

%------------------------------------
%------------------------------------

As an example, we selected five benchmark points labelled $P_{i}^\tau$ across the $1\sigma$ fit region of Figure \ref{fig:SmelliScans}. These scenarios have distinct decay channel magnitudes and are therefore rather illustrative for collider considerations. All information regarding these benchmark points is gathered in Table \ref{tab:BP_tau_iso}.

\begin{figure}[!t]
    \centering
    \includegraphics[scale = 0.7]{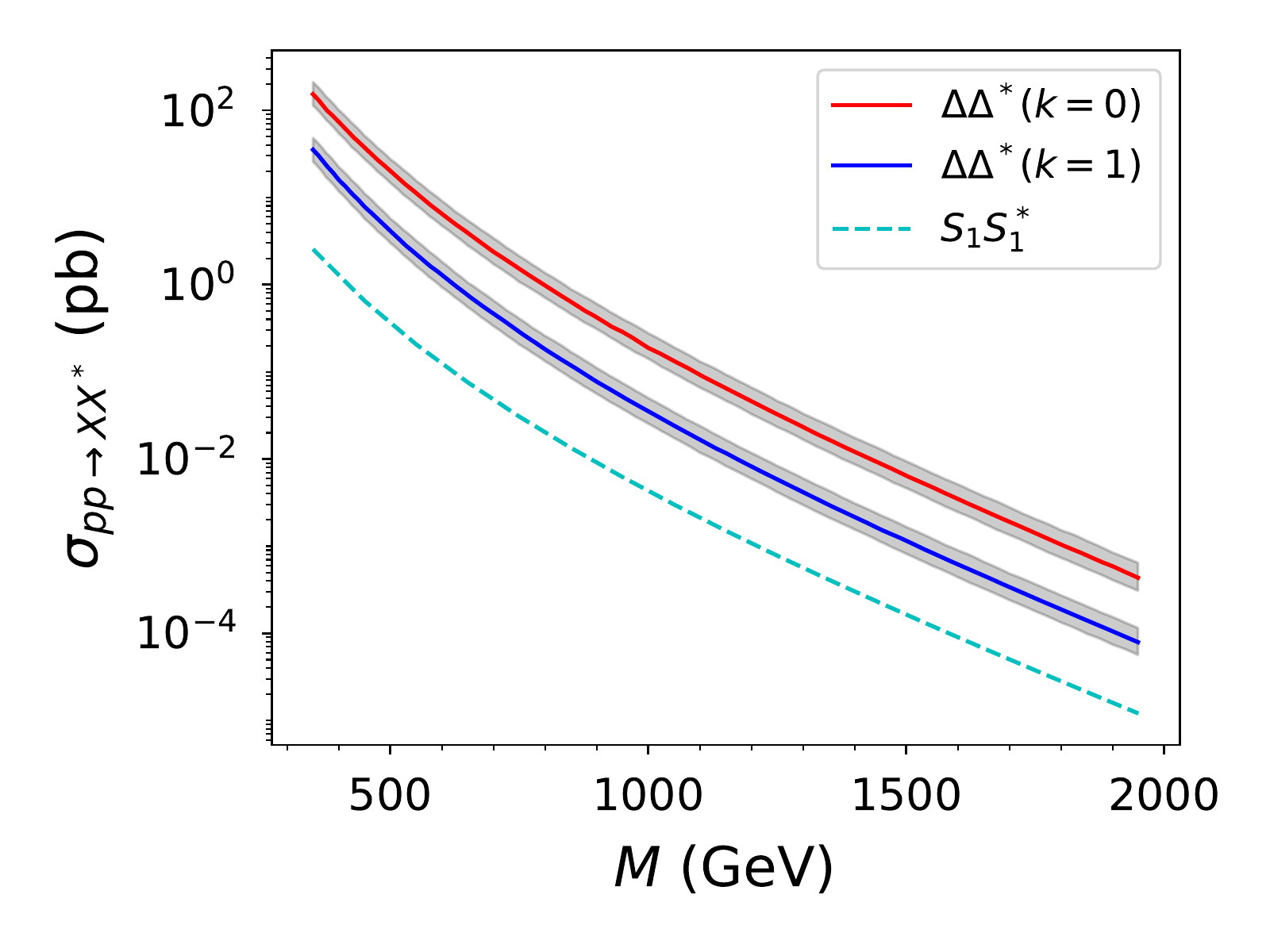}
    \caption{Pair production cross-section for different LQ models. For the vector LQ $\Delta$ we present results for both the $k=0$ and $k=1$ scenarios studied in the text. {Uncertainties are calculated for the vector leptoquark model by varying the scale in \textsc{MadGraph5\_aMC@NLO}~ 2.8.2 from 0.5 to 2, with respect to the central default value, see main text for details.} The scalar model ($S_1$) is presented as an illustration, being the benchmark model employed in most of the ATLAS and CMS studies. }
    \label{fig:pair_production_XS}
\end{figure}

The procedure regarding our collider analysis is as follows:
we first consider the mixed search $\Delta \rightarrow b\tau / t\nu$ investigated in \cite{ATLAS:2021jyv} where we adapt to match the extra opened second-generation quark channel. We then complete the analysis by confronting our model with the implemented LHC searches in {\tt CheckMATE} \cite{Drees:2013wra,Dercks:2016npn}. Of particular interest would be the jets and missing energy searches, since the channel involving jets and neutrinos remains stable around 50\% as discussed in \eqref{eq:BRstructure}. We present results for the two extreme cases for the pair production cross-section: $k = 0$ and $k = 1$, {for which the values are given together with the scalar leptoquark one for comparison in Fig. \ref{fig:pair_production_XS}. }

 %------------------------------------
%------------------------------------

{In this section we compute leading-order cross sections and generate parton level events with the help of 
\textsc{MadGraph5\_aMC@NLO}~ 2.8.2 using an in-house implementation in {\sc \small Feynrules}~\cite{Degrande:2011ua,Alloul:2013bka} of the Lagrangian from Section~\ref{subsec:Lagrangian}. The \texttt{NNPDF30\_lo\_as\_0118}~\cite{NNPDF:2014otw} set of parton distribution functions (PDF) is employed, which we handle through {\sc LHAPDF}~\cite{Buckley:2014ana}. The factorization and renormalization scales are set on an event-by-event basis by \textsc{MadGraph5\_aMC@NLO}, setting the parameters {\tt dynamical\_scale\_choice} and {\tt scalefact} to -1 and 1.0, respectively. We have verified that varying the {\tt scalefact} parameter by a factor of two (to 2.0 and 0.5) impacts the cross-section by -30 \% and +40 \%, respectively. Moreover, this variation is almost independent of the leptoquark mass (scanning in the 0.3-2 TeV range). We have neglected diagrams including a $t$-channel lepton exchange as their contribution is sub-leading with respect to the scale variation~\footnote{For our BP1 benchmark point and with an extreme value of 2 TeV for the LQ mass (currently outside of LHC reach), including the $t$-channel lepton exchange gives a cross section increase of 25 \%. This  agrees with the detailed study in Appendix B of ~\cite{Belanger:2021smw} (albeit  for a scalar LQ model) that found for a 2 TeV LQ an increase of at most 20 \%, restricting oneself to perturbative couplings. }. The technical details for the showering and detector simulation of the parton level events (employed as CheckMATE input) are described in Section~\ref{subsec:CM}.   }

%%%%%%%%%%%%%%%%%%%%%%%%%%%%%%%%%%%%%%%%

\subsection[Reinterpretation of the Mixed $b\tau\,t\nu$ Search]{\boldmath Reinterpretation of the Mixed $b\tau\,t\nu$ Search}

Several searches at the LHC by the ATLAS~\cite{ATLAS:2019qpq,ATLAS:2021jyv} and CMS~\cite{CMS:2018qqq,CMS:2018txo,CMS:2020wzx} collaborations explicitly target up-type vector LQs with $b \tau$ and $t \nu$ final states. Among those, ATLAS in \cite{ATLAS:2021jyv}  (CMS in \cite{CMS:2020wzx}) considers for the first time the possibility of a $b \tau t \nu$ final state (originally pointed out in ~\cite{Gripaios:2010hv} and scrutinized in~\cite{Brooijmans:2020yij,Belanger:2021smw}), where each LQ decayed in a different channel. We informally refer to this as the ``mixed'' channel. Moreover, for the specific case where $\text{BR}(\Delta\rightarrow b\tau)$ and $ \text{BR}(\Delta \rightarrow t\nu)$ are approximately equal, the study of reference~\cite{ATLAS:2021jyv} provides the most stringent constraints, and hence, among the whole suite of LQ studies, we focus on the reinterpretation of this specific final state.

The ATLAS study~\cite{ATLAS:2021jyv} considers LQ decays exclusively to third-generation quarks, and hence the branching ratios into $t \nu$ and $b \tau$ add up to unity. However, in more generic setups additional decay channels can be opened. Following the strategy proposed in \cite{Belanger:2021smw} we reinterpreted the search in this more generic framework, for our vector LQ model. Here we briefly sketch the basics of our procedure, and refer the reader to reference~\cite{Belanger:2021smw} for details. Note that in the following we re-interpret the ATLAS search and not the same final state CMS one \cite{CMS:2020wzx} since the ATLAS search is more sensitive for vector LQs in the parameter space of interest.

The number of events in a signal region is proportional to the pair production cross section $\sigma (p p \to \Delta \Delta) \equiv \sigma$, to the branching ratios for $\Delta \to b \tau$ and $\Delta \to t \nu$ and to the \emph{acceptance} $A$ and \emph{efficiency} $\epsilon$ in the signal regions. The latter two functions are reported in the auxiliary material of ~\cite{ATLAS:2021jyv} for several LQ scenarios, including the minimal coupling ($k=1$) and Yang-Mills ($k=0$)  incarnations of our $\Delta_\mu$ vector case, as a function of the mass and the branching ratio into $b \tau$, where it is assumed that the $\Delta \to t \nu$ channel is the only other channel open. Since this assumption does not hold in general for our benchmark points, the crucial point discussed in Section 3.2.2 of reference~\cite{Belanger:2021smw} is that $A(M, x)$ should be instead $A(M, x')$ where $x= \text{BR}(\Delta \rightarrow b\tau) $ and $x'= \text{BR}(\Delta \rightarrow b\tau) / (\text{BR}(\Delta \rightarrow b\tau)  + \text{BR}(\Delta \rightarrow t \nu) )$; and analogously for the $\epsilon$ functions. The reason behind this choice is the fact that the mixed search should only depend on the relative weight of the $b \tau$ to the $t \nu$ final state. In our case, given by eq.~\eqref{eq:BRstructure}, we see that $x' \approx 0.5$. We do not attempt to reproduce the ATLAS counts in each signal region, since ultimately the final exclusion from ATLAS comes from a combination of signal regions which we can not perform without knowing the relevant correlations, which are regrettably not publicly available. Instead, we will exploit the fact that the ATLAS paper explicitly excludes a mass of 1.50 TeV (1.77 TeV) for $x=0.5$ and $k = 1$ ($k=0$), and we will simply request to have the same events for the exclusion as those obtained for a mass of 1.50 TeV (1.77 TeV) and $\BRthird = 0.5$. Then the excluded value (at $95\%$ C.L.) of $\BRthird$ for a given mass $M$ is given approximately, for the $k=1$ case\footnote{For $k=0$ the same equation holds, replacing 1.5 TeV by 1.77 TeV.},  by 
\begin{equation}
(\BRthird)^{\rm 95} = 0.5 \Biggl(\frac{\sigma(1.5 {\rm TeV}) A(1.5 {\rm TeV}, 0.5) \epsilon (1.5 {\rm TeV}, 0.5) }{\sigma(M) A(M, 0.5) \epsilon (M, 0.5)} \Biggr) \, .     
\end{equation}
While the numerical impact of abandoning this approximation is small, along this work we have nonetheless evaluated the corresponding functions at the corresponding value of $x$, including the finite top mass effects in the $\Delta$ branching ratios.

\begin{figure}[t]
    \centering
    \includegraphics[scale = 0.7]{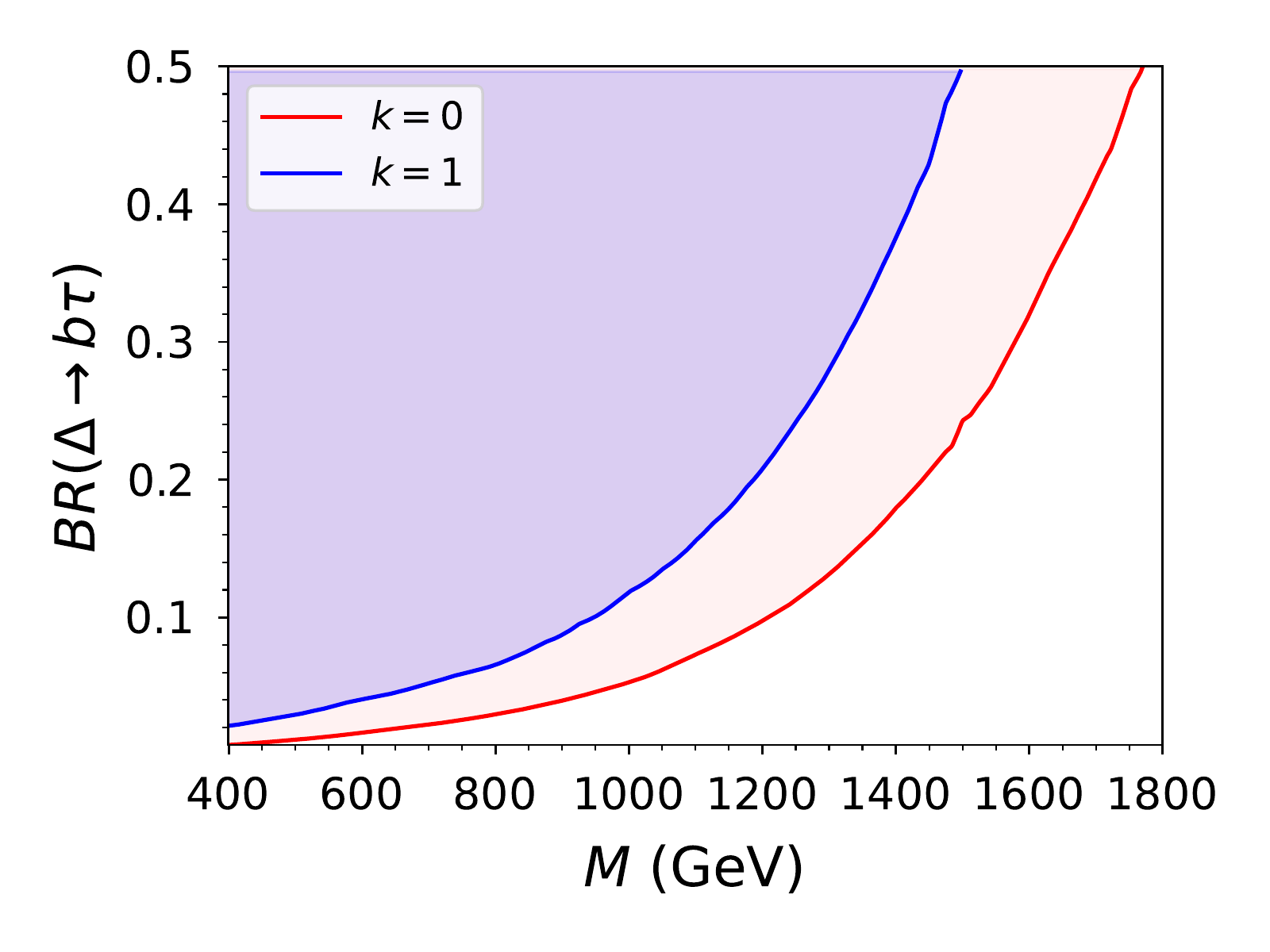}
    \caption{Reinterpretation from \cite{ATLAS:2021jyv}. The main assumptions are $\text{BR}(\Delta\rightarrow b\tau) = \text{BR}(\Delta\rightarrow t\nu)$ and that the opened second generation decay is invisible to the analysis, and hence only affects the fiducial cross-section $2 \times \sigma(pp\rightarrow\Delta \Delta^*) \times (\BRthird)^2$. }
    \label{fig:mixed_LQ_search_rescaled}
\end{figure}

 Following the procedure described above, and using our results for the production cross-section as a function of the mass, given in Figure \ref{fig:pair_production_XS}, it is possible to extend the exclusion given by ATLAS to include the effect of the additional open channels by varying \BRthird instead, which can be matched to the ratio $\lambda_{b\tau}/\lambda_{s\tau}$. The result is presented in Figure \ref{fig:mixed_LQ_search_rescaled}. We observe that, as expected, the excluded mass drops significantly as the branching ratio into the third generation quarks decreases. The two extreme cases from the best fits presented in Section \ref{sec:SCANS} are reached for BP1 and BP5, for which we exclude masses close to 600 GeV (900 GeV) and 1500 GeV (1750 GeV) for $k = 1$ ($k = 0$), respectively.

%----------------------
%----------------------

%%%%%%%%%%%%%%%%%%%%%%%%%%%%%
\subsection{{CheckMATE} Exclusion Limits}
\label{subsec:CM}

As discussed previously, we expect searches involving multi-jets and missing energy to complete the exclusion limits provided by the reinterpretation of the mixed channel. Indeed, at parton level, the branching ratio of the LQ decay channel involving a neutrino ($t\nu + c\nu$) remains stable around 50\% across the coupling parameter space as only the left-handed coupling is present. Decay channels involving either one or two neutrinos will therefore be probed by these searches.

\begin{figure}[t]
    \centering
    \includegraphics[scale = 0.7]{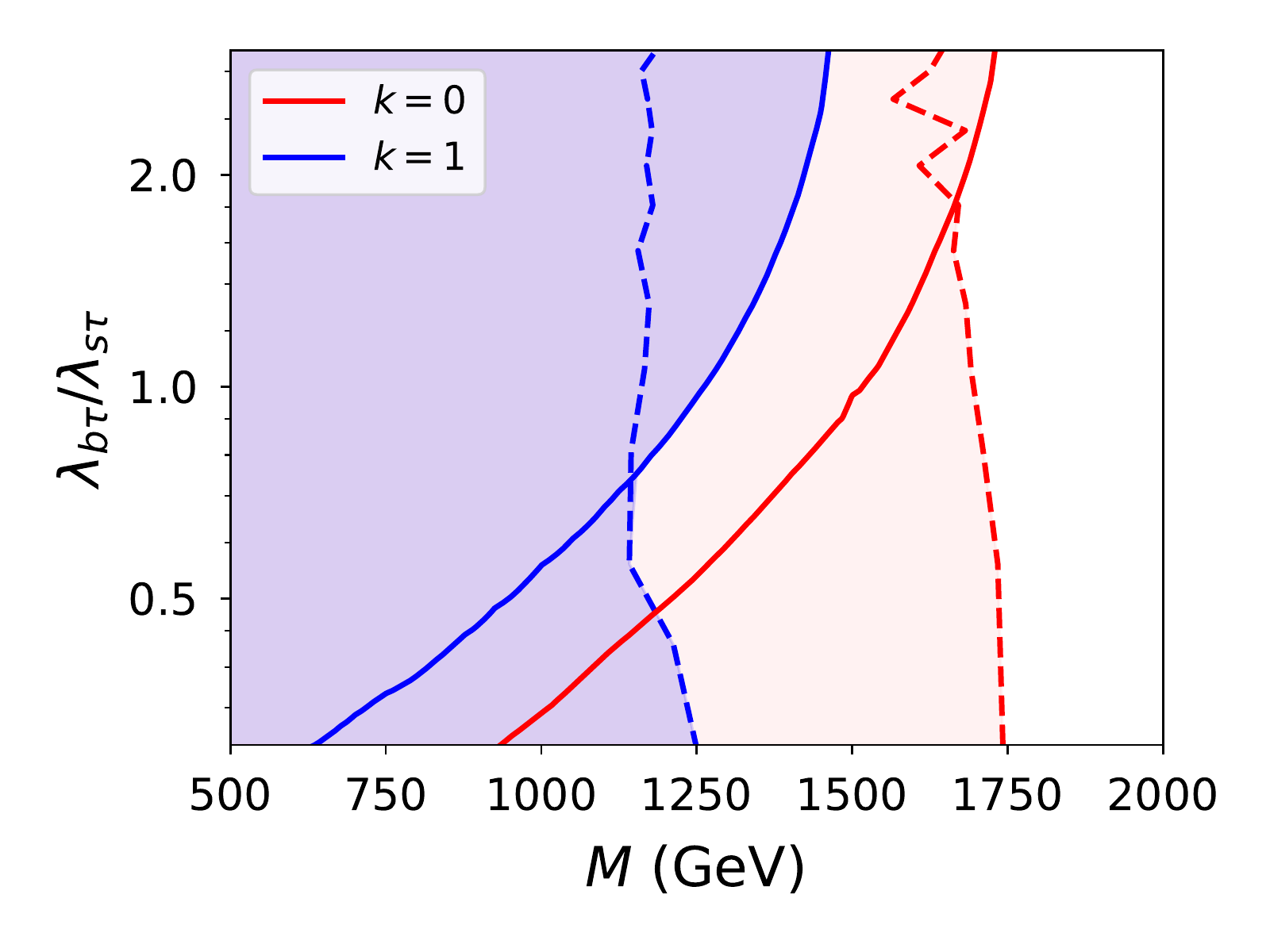}
    \caption{Combined results from {\tt CheckMATE} $nj + \slashed{E}_T$ and mixed-channel exclusion limits. Solid lines correspond to the mixed channel while the dashed ones are the results from {\tt CheckMATE}. Note that the y-axis has a base 2 logarithmic scale.}
    \label{fig:Combined_exclusion}
\end{figure}

We perform a parameter scan in the plane $\left( M, \lambda_{b\tau}/\lambda_{s\tau} \right)$ as only the coupling ratio has an impact on the collider phenomenology.  For each parameter point, we generate 10,000 events at parton level using {\tt MadGraph 5 2.8.2} \cite{Alwall:2014hca} which are then passed through {\tt Pythia 8} \cite{Sjostrand:2014zea} for showering and hadronization. Finally, {\tt Delphes 3.4.2} \cite{deFavereau:2013fsa} and its official ATLAS card are used for fast detector simulation. The full parameter scan is composed of two grids for the two cases $k=0, 1$. For $k=0$ (resp. $k = 1$), the grid is composed of twelve points  $\in [0.31, 3.06]$ (resp. plus two extra points at 0.12, 0.41 needed for better precision) along the $\lambda_{b\tau}/\lambda_{s\tau}$ axis and five (resp. six) points $\in [1675, 1875]$ GeV (resp. $\in [1000, 1250]$ GeV) along the $M$ axis. Using {\tt CheckMATE} v 2.0.34 \cite{Dercks:2016npn} \footnote{{\tt CheckMATE} internally makes use of the anti-$k_t$ algorithm \cite{Cacciari:2008gp} implemented in {\tt FastJet} \cite{Cacciari:2011ma}}, we compute for each point in the parameter space the $r_{95}$ value \cite{Read:2002hq}. This quantity is defined for a given signal region, in a specific experimental analysis, as the ratio between a) the number of predicted signal events and b) the 95\% C.L. exclusion limit provided by the collaboration. Therefore, a point in parameter space is considered excluded if in any of the channels $r_{95} > 1$. Furthermore, {\tt CheckMate} provides together with the $r_{95}$ value, the corresponding most sensitive channel. Using the interpolation function {\tt interp1d} provided by the {\tt scipy} python package, we obtained the $95\%$ C.L. exclusion limit from {\tt CheckMATE} reached for $r_{95} = 1$. 
The results are shown in Fig. \ref{fig:Combined_exclusion} together with the mixed search, adapted to the $\left( M, \lambda_{b\tau}/\lambda_{s\tau} \right)$ plane using the expression for the branching ratios \eqref{eq:simplifiedBR}.

The exclusion is dominated by analyses that involve large missing energy. In particular, the $ nj+\slashed{E}_T$ search released in \cite{ATLAS:2019vcq} is responsible for most of the parameter space exclusion. On the other hand, the searches for sleptons in $n\ell + \slashed{E}_T$ or more specifically staus in $n\tau + \slashed{E}_T$ \cite{ATLAS:2019gti} give the dominant contribution for a smaller set of points. All relevant analyses have a luminosity of 139 fb$^{-1}$.

%------------------------------
%------------------------------
We note that the limits depend only weakly on the coupling ratio with the excluded values for $k = 0$ ($k = 1$) close to $\sim 1700$ GeV ($\sim 1150$ GeV), reproducing our expectations from the parton-level consideration where the neutrino channels sum up to $50\%$ independently of the coupling ratio.

 Being conservative by assuming that the efficiency of future searches will remain stable and therefore simply rescaling by the luminosity, one can estimate prospects for HL-LHC for 3000 fb$^{-1}$, which should exclude masses of order 1400 GeV (2000 GeV) for $k = 1$ ($k = 0$). However, due to the improvement of the analysis techniques (for example in \cite{ATLAS:2020syg} compared to \cite{ATLAS:2017mjy}), one can be more optimistic and expect even higher masses to be reached.  We then see that both classes of studies (searches for LQs and missing energy searches) complement each other and that enhancing the sensitivity of these studies is of crucial importance towards shedding light on the mechanism(s) behind the flavour anomalies.

%%%%%%%%%%%%%%%%%%%%%%%%%%%%%%%%%%%%%%%%%%%%%%%%%%%%%%%%%%%%
\section{Summary and Outlook}
\label{sec:CONCLUDE}
We have studied the phenomenological consequences of a subset of Simplified Models of Flavourful Leptoquarks \cite{deMedeirosVarzielas:2019lgb,Bernigaud:2019bfy,Bernigaud:2020wvn} that only couple to a single generation of charged leptons, with a central focus on the $\left({\bf{3}}, {\bf{1}}, 2/3 \right)$ vector leptoquark $\Delta_\mu$ coupling quarks to $\tau$ leptons.  Lepton isolation patterns/models are readily motivated by ultraviolet flavour-symmetry breaking, and are amongst the most minimal scenarios that can explain experimental anomalies in the lepton-flavour-universality observables $\mathcal{R}_{D^{(\star)},K^{(\star)}}$, depending on the actual tree-level lepton couplings allowed.  We reviewed their theoretical motivation and used {\tt{smelli}} \cite{Aebischer:2018iyb} to scan over the global parameter space favoured by low-energy flavour data when couplings to electrons, muons, or taus are permitted.  This analysis revealed that all three patterns are favoured over SM physics alone, and we presented $1\sigma$ and $2\sigma$ global likelihood contours in the two-parameter spaces of \eqref{eq:yukeisolation}, along with likelihood contours in the same space, but coming from $\mathcal{R}_{D^{(\star)},K^{(\star)}}$ data alone. Upon focusing on the $\tau-$isolation pattern, we then studied the decay signatures of $\Delta_\mu$ at the LHC. 

Based on our parameter-space scans, we have defined a series of benchmark points to illustrate the main phenomenological features, where the branching fraction into the third generation ranges from a few percent up to almost 100 \%. We have confronted the parameter space of our model with two different classes of LHC searches, namely i) direct searches for leptoquarks and ii) missing energy searches. Given that our setup imposes approximately equal branching ratios into $t \nu$ and $b \tau$, it turns out that the mixed ATLAS search looking for $b \tau t \nu$ provides the strongest constraints among the set of direct leptoquark searches. We carried out the reinterpretation of this search for our vector leptoquark models, where the sensitivity would obviously depend on the branching ratio into the third generation.  If no additional channels are open, these searches can constrain a leptoquark mass up to 1500 (1770) GeV for the $k=1$ ($k=0$) scenarios, while if instead the branching fraction into the third generation were to be 5\%, these upper limits would be reduced to approximately 700 (1000) GeV. Regarding the missing energy searches, since our setup also imposes that our leptoquark decays 50\% of the time in a channel with one neutrino, it is not surprising that the mass reach is pretty insensitive to the third generation branching fraction, being about 1150 (1700) GeV for the $k=1$ ($k=0$) case. The complementarity between both types of searches is interesting.

Last but not least, we would like to point out that it would be desirable to expand the set of direct leptoquark searches to specifically target decays into the second generation. While experimentally challenging, it is clear that since the flavour anomalies are directly related to the second generation, searches including $c$ quarks, e.g. $c \nu c \nu$ or $b \tau c \nu$ would have an important impact on  both the discovery and the characterization of leptoquark models. While outside the scope of this work, it would also be desirable to study in similar fashion how future colliders can fully probe the vector leptoquark scenario.

%%%%%%%%%%%%%%%%%%%%%%%%%%%%%%%%%%%%%%%%%%%%%%%%%%%%%%%%%%%%
\section*{Acknowledgements}
We thank Peter Stangl and Jason Aebischer for very helpful discussions regarding {\tt{smelli}}. JB and MB are supported by the Deutsche Forschungsgemeinschaft (DFG, German Research Foundation) under grant  396021762 -- TRR 257.
IdMV acknowledges funding from Funda\c{c}\~{a}o para a Ci\^{e}ncia e a Tecnologia (FCT, Portugal) through the
contract IF/00816/2015 and through the projects CFTP-FCT Unit 777
(UIDB/00777/2020 and UIDP/00777/2020), PTDC/FIS-PAR/29436/2017 and CERN/FIS-PAR/0008/2019, which are partially funded through POCTI (FEDER), COMPETE, QREN and EU. 
JT gratefully acknowledges funding from the European Union’s Horizon 2020 research and innovation programme under the Marie Sk\l{}odowska-Curie grant agreement No. 101022203, as well as prior support from the Villum Fund (project No. 00010102). JZ is supported by the {\it Generalitat Valenciana} (Spain) through the {\it plan GenT} program (CIDEGENT/2019/068), by the Spanish Government (Agencia Estatal de
Investigación) and ERDF funds from European Commission \\ (MCIN/AEI/10.13039/501100011033, Grant No. PID2020-114473GB-I00).
%%%%%%%%%%%%%%%%%%%%%%%%%%%%%%%%%%%%%%%%%%%%%%%%%%%%%%%%%%%%


\begin{thebibliography}{9}
%\cite{Lees:2012xj}
\bibitem{Lees:2012xj}
J.~P.~Lees \textit{et al.} [BaBar],
%``Evidence for an excess of $\bar{B} \to D^{(*)} \tau^-\bar{\nu}_\tau$ decays,''
Phys. Rev. Lett. \textbf{109} (2012), 101802
doi:10.1103/PhysRevLett.109.101802
[arXiv:1205.5442 [hep-ex]].
%979 citations counted in INSPIRE as of 21 Dec 2021

%\cite{Lees:2013uzd}
\bibitem{Lees:2013uzd}
J.~P.~Lees \textit{et al.} [BaBar],
%``Measurement of an Excess of $\bar{B} \to D^{(*)}\tau^- \bar{\nu}_\tau$ Decays and Implications for Charged Higgs Bosons,''
Phys. Rev. D \textbf{88} (2013) no.7, 072012
doi:10.1103/PhysRevD.88.072012
[arXiv:1303.0571 [hep-ex]].
%829 citations counted in INSPIRE as of 21 Dec 2021

%\cite{Huschle:2015rga}
\bibitem{Huschle:2015rga}
M.~Huschle \textit{et al.} [Belle],
%``Measurement of the branching ratio of $\bar{B} \to D^{(\ast)} \tau^- \bar{\nu}_\tau$ relative to $\bar{B} \to D^{(\ast)} \ell^- \bar{\nu}_\ell$ decays with hadronic tagging at Belle,''
Phys. Rev. D \textbf{92} (2015) no.7, 072014
doi:10.1103/PhysRevD.92.072014
[arXiv:1507.03233 [hep-ex]].
%763 citations counted in INSPIRE as of 21 Dec 2021

%\cite{Hirose:2016wfn}
\bibitem{Hirose:2016wfn}
S.~Hirose \textit{et al.} [Belle],
%``Measurement of the $\tau$ lepton polarization and $R(D^*)$ in the decay $\bar{B} \to D^* \tau^- \bar{\nu}_\tau$,''
Phys. Rev. Lett. \textbf{118} (2017) no.21, 211801
doi:10.1103/PhysRevLett.118.211801
[arXiv:1612.00529 [hep-ex]].
%494 citations counted in INSPIRE as of 21 Dec 2021

%\cite{Belle:2016ure}
\bibitem{Belle:2016ure}
Y.~Sato \textit{et al.} [Belle],
%``Measurement of the branching ratio of $\bar{B}^0 \rightarrow D^{*+} \tau^- \bar{\nu}_{\tau}$ relative to $\bar{B}^0 \rightarrow D^{*+} \ell^- \bar{\nu}_{\ell}$ decays with a semileptonic tagging method,''
Phys. Rev. D \textbf{94} (2016) no.7, 072007
doi:10.1103/PhysRevD.94.072007
[arXiv:1607.07923 [hep-ex]].
%354 citations counted in INSPIRE as of 21 Dec 2021

%\cite{Belle:2016dyj}
\bibitem{Belle:2016dyj}
S.~Hirose \textit{et al.} [Belle],
%``Measurement of the $\tau$ lepton polarization and $R(D^*)$ in the decay $\bar{B} \to D^* \tau^- \bar{\nu}_\tau$,''
Phys. Rev. Lett. \textbf{118} (2017) no.21, 211801
doi:10.1103/PhysRevLett.118.211801
[arXiv:1612.00529 [hep-ex]].
%494 citations counted in INSPIRE as of 21 Dec 2021

%\cite{Belle:2019gij}
\bibitem{Belle:2019gij}
A.~Abdesselam \textit{et al.} [Belle],
%``Measurement of $\mathcal{R}(D)$ and $\mathcal{R}(D^{\ast})$ with a semileptonic tagging method,''
[arXiv:1904.08794 [hep-ex]].
%202 citations counted in INSPIRE as of 21 Dec 2021

%\cite{Aaij:2015yra}
\bibitem{Aaij:2015yra}
R.~Aaij \textit{et al.} [LHCb],
%``Measurement of the ratio of branching fractions $\mathcal{B}(\bar{B}^0 \to D^{*+}\tau^{-}\bar{\nu}_{\tau})/\mathcal{B}(\bar{B}^0 \to D^{*+}\mu^{-}\bar{\nu}_{\mu})$,''
Phys. Rev. Lett. \textbf{115} (2015) no.11, 111803
[erratum: Phys. Rev. Lett. \textbf{115} (2015) no.15, 159901]
doi:10.1103/PhysRevLett.115.111803
[arXiv:1506.08614 [hep-ex]].
%960 citations counted in INSPIRE as of 22 Dec 2021

%\cite{LHCb:2017smo}
\bibitem{LHCb:2017smo}
R.~Aaij \textit{et al.} [LHCb],
%``Measurement of the ratio of the $B^0 \to D^{*-} \tau^+ \nu_{\tau}$ and $B^0 \to D^{*-} \mu^+ \nu_{\mu}$ branching fractions using three-prong $\tau$-lepton decays,''
Phys. Rev. Lett. \textbf{120} (2018) no.17, 171802
doi:10.1103/PhysRevLett.120.171802
[arXiv:1708.08856 [hep-ex]].
%335 citations counted in INSPIRE as of 22 Dec 2021

%\cite{Aaij:2017deq}
\bibitem{Aaij:2017deq}
R.~Aaij \textit{et al.} [LHCb],
%``Test of Lepton Flavor Universality by the measurement of the $B^0 \to D^{*-} \tau^+ \nu_{\tau}$ branching fraction using three-prong $\tau$ decays,''
Phys. Rev. D \textbf{97} (2018) no.7, 072013
doi:10.1103/PhysRevD.97.072013
[arXiv:1711.02505 [hep-ex]].
%281 citations counted in INSPIRE as of 22 Dec 2021

%\cite{HFLAV:2019otj}
\bibitem{HFLAV:2019otj}
Y.~S.~Amhis \textit{et al.} [HFLAV],
%``Averages of b-hadron, c-hadron, and $\tau $-lepton properties as of 2018,''
Eur. Phys. J. C \textbf{81} (2021) no.3, 226
doi:10.1140/epjc/s10052-020-8156-7
[arXiv:1909.12524 [hep-ex]].
%459 citations counted in INSPIRE as of 22 Dec 2021

%\cite{Crivellin:2012ye}
\bibitem{Crivellin:2012ye}
A.~Crivellin, C.~Greub and A.~Kokulu,
%``Explaining $B\to D\tau\nu$, $B\to D^*\tau\nu$ and $B\to \tau\nu$ in a 2HDM of type III,''
Phys. Rev. D \textbf{86} (2012), 054014
doi:10.1103/PhysRevD.86.054014
[arXiv:1206.2634 [hep-ph]].
%288 citations counted in INSPIRE as of 05 Dec 2021

%\cite{Crivellin:2013wna}
\bibitem{Crivellin:2013wna}
A.~Crivellin, A.~Kokulu and C.~Greub,
%``Flavor-phenomenology of two-Higgs-doublet models with generic Yukawa structure,''
Phys. Rev. D \textbf{87} (2013) no.9, 094031
doi:10.1103/PhysRevD.87.094031
[arXiv:1303.5877 [hep-ph]].
%263 citations counted in INSPIRE as of 03 Dec 2021

%\cite{Celis:2012dk}
\bibitem{Celis:2012dk}
A.~Celis, M.~Jung, X.~Q.~Li and A.~Pich,
%``Sensitivity to charged scalars in $\boldsymbol{B\to D^{(*)}\tau\nu_\tau}$ and $\boldsymbol{B\to\tau\nu_\tau}$ decays,''
JHEP \textbf{01} (2013), 054
doi:10.1007/JHEP01(2013)054
[arXiv:1210.8443 [hep-ph]].
%251 citations counted in INSPIRE as of 05 Dec 2021

%\cite{Ko:2012sv}
\bibitem{Ko:2012sv}
P.~Ko, Y.~Omura and C.~Yu,
%``$B \to D^(*) \tau \nu$ and $B \to \tau \nu$ in chiral $U(1)^\prime$ models with flavored multi Higgs doublets,''
JHEP \textbf{03} (2013), 151
doi:10.1007/JHEP03(2013)151
[arXiv:1212.4607 [hep-ph]].
%73 citations counted in INSPIRE as of 07 Dec 2021

%\cite{Crivellin:2015hha}
\bibitem{Crivellin:2015hha}
A.~Crivellin, J.~Heeck and P.~Stoffer,
%``A perturbed lepton-specific two-Higgs-doublet model facing experimental hints for physics beyond the Standard Model,''
Phys. Rev. Lett. \textbf{116} (2016) no.8, 081801
doi:10.1103/PhysRevLett.116.081801
[arXiv:1507.07567 [hep-ph]].
%186 citations counted in INSPIRE as of 03 Dec 2021

%\cite{He:2012zp}
\bibitem{He:2012zp}
X.~G.~He and G.~Valencia,
%``$B$ decays with $\tau$ leptons in nonuniversal left-right models,''
Phys. Rev. D \textbf{87} (2013) no.1, 014014
doi:10.1103/PhysRevD.87.014014
[arXiv:1211.0348 [hep-ph]].
%80 citations counted in INSPIRE as of 03 Dec 2021

%\cite{Greljo:2015mma}
\bibitem{Greljo:2015mma}
A.~Greljo, G.~Isidori and D.~Marzocca,
%``On the breaking of Lepton Flavor Universality in B decays,''
JHEP \textbf{07} (2015), 142
doi:10.1007/JHEP07(2015)142
[arXiv:1506.01705 [hep-ph]].
%272 citations counted in INSPIRE as of 15 Dec 2021

%\cite{Boucenna:2016wpr}
\bibitem{Boucenna:2016wpr}
S.~M.~Boucenna, A.~Celis, J.~Fuentes-Martin, A.~Vicente and J.~Virto,
%``Non-abelian gauge extensions for B-decay anomalies,''
Phys. Lett. B \textbf{760} (2016), 214-219
doi:10.1016/j.physletb.2016.06.067
[arXiv:1604.03088 [hep-ph]].
%169 citations counted in INSPIRE as of 15 Dec 2021

%\cite{He:2017bft}
\bibitem{He:2017bft}
X.~G.~He and G.~Valencia,
%``Lepton universality violation and right-handed currents in $b \to c \tau \nu$,''
Phys. Lett. B \textbf{779} (2018), 52-57
doi:10.1016/j.physletb.2018.01.073
[arXiv:1711.09525 [hep-ph]].
%44 citations counted in INSPIRE as of 03 Dec 2021

%\cite{Alonso:2015sja}
\bibitem{Alonso:2015sja}
R.~Alonso, B.~Grinstein and J.~Martin Camalich,
%``Lepton universality violation and lepton flavor conservation in $B$-meson decays,''
JHEP \textbf{10} (2015), 184
doi:10.1007/JHEP10(2015)184
[arXiv:1505.05164 [hep-ph]].
%339 citations counted in INSPIRE as of 05 Dec 2021

%\cite{Calibbi:2015kma}
\bibitem{Calibbi:2015kma}
L.~Calibbi, A.~Crivellin and T.~Ota,
%``Effective Field Theory Approach to $b\to s\ell\ell^{(')}$, $B\to  K^{(*)}\nu\overline{\nu}$ and $B\to D^{(*)}\tau\nu$ with Third  Generation Couplings,''
Phys. Rev. Lett. \textbf{115} (2015), 181801
doi:10.1103/PhysRevLett.115.181801
[arXiv:1506.02661 [hep-ph]].
%296 citations counted in INSPIRE as of 20 Dec 2021

%\cite{Fajfer:2015ycq}
\bibitem{Fajfer:2015ycq}
S.~Fajfer and N.~Ko\v{s}nik,
%``Vector leptoquark resolution of $R_K$ and $R_{D^{(*)}}$ puzzles,''
Phys. Lett. B \textbf{755} (2016), 270-274
doi:10.1016/j.physletb.2016.02.018
[arXiv:1511.06024 [hep-ph]].
%250 citations counted in INSPIRE as of 05 Dec 2021

%\cite{Barbieri:2015yvd}
\bibitem{Barbieri:2015yvd}
R.~Barbieri, G.~Isidori, A.~Pattori and F.~Senia,
%``Anomalies in $B$-decays and $U(2)$ flavour symmetry,''
Eur. Phys. J. C \textbf{76} (2016) no.2, 67
doi:10.1140/epjc/s10052-016-3905-3
[arXiv:1512.01560 [hep-ph]].
%256 citations counted in INSPIRE as of 03 Dec 2021

%\cite{Barbieri:2016las}
\bibitem{Barbieri:2016las}
R.~Barbieri, C.~W.~Murphy and F.~Senia,
%``B-decay Anomalies in a Composite Leptoquark Model,''
Eur. Phys. J. C \textbf{77} (2017) no.1, 8
doi:10.1140/epjc/s10052-016-4578-7
[arXiv:1611.04930 [hep-ph]].
%167 citations counted in INSPIRE as of 03 Dec 2021

%\cite{Hiller:2016kry}
\bibitem{Hiller:2016kry}
G.~Hiller, D.~Loose and K.~Sch\"onwald,
%``Leptoquark Flavor Patterns \& B Decay Anomalies,''
JHEP \textbf{12} (2016), 027
doi:10.1007/JHEP12(2016)027
[arXiv:1609.08895 [hep-ph]].
%147 citations counted in INSPIRE as of 20 Dec 2021

%\cite{Deshpande:2012rr}
\bibitem{Deshpande:2012rr}
N.~G.~Deshpande and A.~Menon,
%``Hints of R-parity violation in B decays into $\tau \nu$,''
JHEP \textbf{01} (2013), 025
doi:10.1007/JHEP01(2013)025
[arXiv:1208.4134 [hep-ph]].
%84 citations counted in INSPIRE as of 03 Dec 2021

%\cite{Tanaka:2012nw}
\bibitem{Tanaka:2012nw}
M.~Tanaka and R.~Watanabe,
%``New physics in the weak interaction of $\bar B\to D^{(*)}\tau\bar\nu$,''
Phys. Rev. D \textbf{87} (2013) no.3, 034028
doi:10.1103/PhysRevD.87.034028
[arXiv:1212.1878 [hep-ph]].
%270 citations counted in INSPIRE as of 05 Dec 2021

%\cite{Sakaki:2013bfa}
\bibitem{Sakaki:2013bfa}
Y.~Sakaki, M.~Tanaka, A.~Tayduganov and R.~Watanabe,
%``Testing leptoquark models in $\bar B \to D^{(*)} \tau \bar\nu$,''
Phys. Rev. D \textbf{88} (2013) no.9, 094012
doi:10.1103/PhysRevD.88.094012
[arXiv:1309.0301 [hep-ph]].
%291 citations counted in INSPIRE as of 10 Dec 2021

%\cite{Freytsis:2015qca}
\bibitem{Freytsis:2015qca}
M.~Freytsis, Z.~Ligeti and J.~T.~Ruderman,
%``Flavor models for $\bar{B} \to D^{(*)} \tau \bar{\nu}$,''
Phys. Rev. D \textbf{92} (2015) no.5, 054018
doi:10.1103/PhysRevD.92.054018
[arXiv:1506.08896 [hep-ph]].
%287 citations counted in INSPIRE as of 21 Dec 2021

%\cite{Bauer:2015knc}
\bibitem{Bauer:2015knc}
M.~Bauer and M.~Neubert,
%``Minimal Leptoquark Explanation for the $R_{D^{(*)}}$ , $R_K$ , and $(g-2)_\mu$ Anomalies,''
Phys. Rev. Lett. \textbf{116} (2016) no.14, 141802
doi:10.1103/PhysRevLett.116.141802
[arXiv:1511.01900 [hep-ph]].
%402 citations counted in INSPIRE as of 10 Dec 2021

%\cite{Becirevic:2016yqi}
\bibitem{Becirevic:2016yqi}
D.~Be\v{c}irevi\'c, S.~Fajfer, N.~Ko\v{s}nik and O.~Sumensari,
%``Leptoquark model to explain the $B$-physics anomalies, $R_K$ and $R_D$,''
Phys. Rev. D \textbf{94} (2016) no.11, 115021
doi:10.1103/PhysRevD.94.115021
[arXiv:1608.08501 [hep-ph]].
%243 citations counted in INSPIRE as of 21 Dec 2021

%\cite{Becirevic:2018afm}
\bibitem{Becirevic:2018afm}
D.~Be\v{c}irevi\'c, I.~Dor\v{s}ner, S.~Fajfer, N.~Ko\v{s}nik, D.~A.~Faroughy and O.~Sumensari,
%``Scalar leptoquarks from grand unified theories to accommodate the $B$-physics anomalies,''
Phys. Rev. D \textbf{98} (2018) no.5, 055003
doi:10.1103/PhysRevD.98.055003
[arXiv:1806.05689 [hep-ph]].
%152 citations counted in INSPIRE as of 10 Dec 2021

%\cite{deMedeirosVarzielas:2019lgb}
\bibitem{deMedeirosVarzielas:2019lgb}
I.~de Medeiros Varzielas and J.~Talbert,
%``Simplified Models of Flavourful Leptoquarks,''
Eur. Phys. J. C \textbf{79} (2019) no.6, 536
doi:10.1140/epjc/s10052-019-7047-2
[arXiv:1901.10484 [hep-ph]].
%22 citations counted in INSPIRE as of 03 Dec 2021

%\cite{Bernigaud:2019bfy}
\bibitem{Bernigaud:2019bfy}
J.~Bernigaud, I.~de Medeiros Varzielas and J.~Talbert,
%``Finite Family Groups for Fermionic and Leptoquark Mixing Patterns,''
JHEP \textbf{01} (2020), 194
doi:10.1007/JHEP01(2020)194
[arXiv:1906.11270 [hep-ph]].
%13 citations counted in INSPIRE as of 03 Dec 2021

%\cite{Bernigaud:2020wvn}
\bibitem{Bernigaud:2020wvn}
J.~Bernigaud, I.~de Medeiros Varzielas and J.~Talbert,
%``Reconstructing Effective Lagrangians Embedding Residual Family Symmetries,''
Eur. Phys. J. C \textbf{81} (2021) no.1, 65
doi:10.1140/epjc/s10052-021-08882-7
[arXiv:2005.12293 [hep-ph]].
%1 citations counted in INSPIRE as of 03 Dec 2021

%\cite{Aebischer:2019mlg}
\bibitem{Aebischer:2019mlg}
J.~Aebischer, W.~Altmannshofer, D.~Guadagnoli, M.~Reboud, P.~Stangl and D.~M.~Straub,
%``$B$-decay discrepancies after Moriond 2019,''
Eur. Phys. J. C \textbf{80} (2020) no.3, 252
doi:10.1140/epjc/s10052-020-7817-x
[arXiv:1903.10434 [hep-ph]].
%234 citations counted in INSPIRE as of 20 Dec 2021

%\cite{Assad:2017iib}
\bibitem{Assad:2017iib}
N.~Assad, B.~Fornal and B.~Grinstein,
%``Baryon Number and Lepton Universality Violation in Leptoquark and Diquark Models,''
Phys. Lett. B \textbf{777} (2018), 324-331
doi:10.1016/j.physletb.2017.12.042
[arXiv:1708.06350 [hep-ph]].
%166 citations counted in INSPIRE as of 05 Jan 2022

%\cite{Pati:1974yy}
\bibitem{Pati:1974yy}
J.~C.~Pati and A.~Salam,
%``Lepton Number as the Fourth Color,''
Phys. Rev. D \textbf{10} (1974), 275-289
[erratum: Phys. Rev. D \textbf{11} (1975), 703-703]
doi:10.1103/PhysRevD.10.275
%5170 citations counted in INSPIRE as of 20 Dec 2021

%\cite{DiLuzio:2017vat}
\bibitem{DiLuzio:2017vat}
L.~Di Luzio, A.~Greljo and M.~Nardecchia,
%``Gauge leptoquark as the origin of B-physics anomalies,''
Phys. Rev. D \textbf{96} (2017) no.11, 115011
doi:10.1103/PhysRevD.96.115011
[arXiv:1708.08450 [hep-ph]].
%205 citations counted in INSPIRE as of 03 Dec 2021

%\cite{Calibbi:2017qbu}
\bibitem{Calibbi:2017qbu}
L.~Calibbi, A.~Crivellin and T.~Li,
%``Model of vector leptoquarks in view of the $B$-physics anomalies,''
Phys. Rev. D \textbf{98} (2018) no.11, 115002
doi:10.1103/PhysRevD.98.115002
[arXiv:1709.00692 [hep-ph]].
%193 citations counted in INSPIRE as of 13 Dec 2021

%\cite{Bordone:2017bld}
\bibitem{Bordone:2017bld}
M.~Bordone, C.~Cornella, J.~Fuentes-Martin and G.~Isidori,
%``A three-site gauge model for flavor hierarchies and flavor anomalies,''
Phys. Lett. B \textbf{779} (2018), 317-323
doi:10.1016/j.physletb.2018.02.011
[arXiv:1712.01368 [hep-ph]].
%179 citations counted in INSPIRE as of 03 Dec 2021

%\cite{Barbieri:2017tuq}
\bibitem{Barbieri:2017tuq}
R.~Barbieri and A.~Tesi,
%``$B$-decay anomalies in Pati-Salam SU(4),''
Eur. Phys. J. C \textbf{78} (2018) no.3, 193
doi:10.1140/epjc/s10052-018-5680-9
[arXiv:1712.06844 [hep-ph]].
%106 citations counted in INSPIRE as of 03 Dec 2021

%\cite{Blanke:2018sro}
\bibitem{Blanke:2018sro}
M.~Blanke and A.~Crivellin,
%``$B$ Meson Anomalies in a Pati-Salam Model within the Randall-Sundrum Background,''
Phys. Rev. Lett. \textbf{121} (2018) no.1, 011801
doi:10.1103/PhysRevLett.121.011801
[arXiv:1801.07256 [hep-ph]].
%182 citations counted in INSPIRE as of 13 Dec 2021

%\cite{Alguero:2021anc}
\bibitem{Alguero:2021anc}
M.~Alguer\'o, B.~Capdevila, S.~Descotes-Genon, J.~Matias and M.~Novoa-Brunet,
%``$\boldsymbol{b\to s\ell\ell}$ global fits after Moriond 2021 results,''
[arXiv:2104.08921 [hep-ph]].
%50 citations counted in INSPIRE as of 21 Dec 2021

%\cite{Altmannshofer:2021qrr}
\bibitem{Altmannshofer:2021qrr}
W.~Altmannshofer and P.~Stangl,
%``New physics in rare B decays after Moriond 2021,''
Eur. Phys. J. C \textbf{81} (2021) no.10, 952
doi:10.1140/epjc/s10052-021-09725-1
[arXiv:2103.13370 [hep-ph]].
%91 citations counted in INSPIRE as of 21 Dec 2021

%\cite{LHCb:2017avl}
\bibitem{LHCb:2017avl}
R.~Aaij \textit{et al.} [LHCb],
%``Test of lepton universality with $B^{0} \rightarrow K^{*0}\ell^{+}\ell^{-}$ decays,''
JHEP \textbf{08} (2017), 055
doi:10.1007/JHEP08(2017)055
[arXiv:1705.05802 [hep-ex]].
%949 citations counted in INSPIRE as of 22 Dec 2021

%\cite{LHCb:2021trn}
\bibitem{LHCb:2021trn}
R.~Aaij \textit{et al.} [LHCb],
%``Test of lepton universality in beauty-quark decays,''
[arXiv:2103.11769 [hep-ex]].
%203 citations counted in INSPIRE as of 22 Dec 2021

%\cite{LHCb:2021lvy}
\bibitem{LHCb:2021lvy}
R.~Aaij \textit{et al.} [LHCb],
%``Tests of lepton universality using $B^0\to K^0_S \ell^+ \ell^-$ and $B^+\to K^{*+} \ell^+ \ell^-$ decays,''
[arXiv:2110.09501 [hep-ex]].
%11 citations counted in INSPIRE as of 22 Dec 2021

%\cite{BELLE:2019xld}
\bibitem{BELLE:2019xld}
S.~Choudhury \textit{et al.} [BELLE],
%``Test of lepton flavor universality and search for lepton flavor violation in $B \rightarrow K\ell \ell$ decays,''
JHEP \textbf{03} (2021), 105
doi:10.1007/JHEP03(2021)105
[arXiv:1908.01848 [hep-ex]].
%107 citations counted in INSPIRE as of 20 Dec 2021

%\cite{Belle:2019oag}
\bibitem{Belle:2019oag}
A.~Abdesselam \textit{et al.} [Belle],
%``Test of Lepton-Flavor Universality in ${B\to K^\ast\ell^+\ell^-}$ Decays at Belle,''
Phys. Rev. Lett. \textbf{126} (2021) no.16, 161801
doi:10.1103/PhysRevLett.126.161801
[arXiv:1904.02440 [hep-ex]].
%202 citations counted in INSPIRE as of 20 Dec 2021

%\cite{Allanach:2017bta}
\bibitem{Allanach:2017bta}
B.~C.~Allanach, B.~Gripaios and T.~You,
%``The case for future hadron colliders from $B \to K^{(*)} \mu^+ \mu^-$  decays,''
JHEP \textbf{03} (2018), 021
doi:10.1007/JHEP03(2018)021
[arXiv:1710.06363 [hep-ph]].
%44 citations counted in INSPIRE as of 10 Dec 2021

%\cite{Garland:2021ghw}
\bibitem{Garland:2021ghw}
B.~Garland, S.~J\"ager, C.~K.~Khosa and S.~Kvedarait\.{e},
%``Probing B-Anomalies via Dimuon Tails at a Future Collider,''
[arXiv:2112.05127 [hep-ph]].
%0 citations counted in INSPIRE as of 16 Dec 2021

%\cite{Asadi:2021gah}
\bibitem{Asadi:2021gah}
P.~Asadi, R.~Capdevilla, C.~Cesarotti and S.~Homiller,
%``Searching for leptoquarks at future muon colliders,''
JHEP \textbf{10} (2021), 182
doi:10.1007/JHEP10(2021)182
[arXiv:2104.05720 [hep-ph]].
%9 citations counted in INSPIRE as of 09 Dec 2021

%\cite{Allanach:2019zfr}
\bibitem{Allanach:2019zfr}
B.~C.~Allanach, T.~Corbett and M.~Madigan,
%``Sensitivity of Future Hadron Colliders to Leptoquark Pair Production in the Di-Muon Di-Jets Channel,''
Eur. Phys. J. C \textbf{80} (2020) no.2, 170
doi:10.1140/epjc/s10052-020-7722-3
[arXiv:1911.04455 [hep-ph]].
%19 citations counted in INSPIRE as of 10 Dec 2021

%\cite{Mandal:2015vfa}
\bibitem{Mandal:2015vfa}
T.~Mandal, S.~Mitra and S.~Seth,
%``Single Productions of Colored Particles at the LHC: An Example with Scalar Leptoquarks,''
JHEP \textbf{07} (2015), 028
doi:10.1007/JHEP07(2015)028
[arXiv:1503.04689 [hep-ph]].
%41 citations counted in INSPIRE as of 03 Dec 2021

%\cite{Bhaskar:2021gsy}
\bibitem{Bhaskar:2021gsy}
A.~Bhaskar, T.~Mandal, S.~Mitra and M.~Sharma,
%``Improving third-generation leptoquark searches with combined signals and boosted top quarks,''
Phys. Rev. D \textbf{104} (2021) no.7, 075037
doi:10.1103/PhysRevD.104.075037
[arXiv:2106.07605 [hep-ph]].
%4 citations counted in INSPIRE as of 09 Dec 2021

%\cite{Aydemir:2019ynb}
\bibitem{Aydemir:2019ynb}
U.~Aydemir, T.~Mandal and S.~Mitra,
%``Addressing the ${\mathbf R_{D^{(*)}}}$ anomalies with an ${\mathbf S_1}$ leptoquark from $\mathbf{SO(10)}$ grand unification,''
Phys. Rev. D \textbf{101} (2020) no.1, 015011
doi:10.1103/PhysRevD.101.015011
[arXiv:1902.08108 [hep-ph]].
%29 citations counted in INSPIRE as of 07 Dec 2021

%\cite{Baker:2019sli}
\bibitem{Baker:2019sli}
M.~J.~Baker, J.~Fuentes-Mart\'\i{}n, G.~Isidori and M.~K\"onig,
%``High- $p_T$ signatures in vector\textendash{}leptoquark models,''
Eur. Phys. J. C \textbf{79} (2019) no.4, 334
doi:10.1140/epjc/s10052-019-6853-x
[arXiv:1901.10480 [hep-ph]].
%61 citations counted in INSPIRE as of 13 Dec 2021

%\cite{Iguro:2020keo}
\bibitem{Iguro:2020keo}
S.~Iguro, M.~Takeuchi and R.~Watanabe,
%``Testing leptoquark/EFT in ${\bar{B}} \rightarrow {D^{(*)}}l{\bar{\nu }}$ at the LHC,''
Eur. Phys. J. C \textbf{81} (2021) no.5, 406
doi:10.1140/epjc/s10052-021-09125-5
[arXiv:2011.02486 [hep-ph]].
%12 citations counted in INSPIRE as of 05 Jan 2022

%\cite{Endo:2021lhi}
\bibitem{Endo:2021lhi}
M.~Endo, S.~Iguro, T.~Kitahara, M.~Takeuchi and R.~Watanabe,
%``Non-resonant new physics search at the LHC for the $b \to c \tau \nu$ anomalies,''
[arXiv:2111.04748 [hep-ph]].
%0 citations counted in INSPIRE as of 09 Dec 2021

%\cite{Cornella:2021sby}
\bibitem{Cornella:2021sby}
C.~Cornella, D.~A.~Faroughy, J.~Fuentes-Martin, G.~Isidori and M.~Neubert,
%``Reading the footprints of the B-meson flavor anomalies,''
JHEP \textbf{08} (2021), 050
doi:10.1007/JHEP08(2021)050
[arXiv:2103.16558 [hep-ph]].
%52 citations counted in INSPIRE as of 22 Dec 2021

%\cite{Angelescu:2018tyl}
\bibitem{Angelescu:2018tyl}
A.~Angelescu, D.~Be\v{c}irevi\'c, D.~A.~Faroughy and O.~Sumensari,
%``Closing the window on single leptoquark solutions to the $B$-physics anomalies,''
JHEP \textbf{10} (2018), 183
doi:10.1007/JHEP10(2018)183
[arXiv:1808.08179 [hep-ph]].
%200 citations counted in INSPIRE as of 20 Dec 2021

%\cite{Feruglio:2018fxo}
\bibitem{Feruglio:2018fxo}
F.~Feruglio, P.~Paradisi and O.~Sumensari,
%``Implications of scalar and tensor explanations of $R_{D^{(\ast)}}$,''
JHEP \textbf{11} (2018), 191
doi:10.1007/JHEP11(2018)191
[arXiv:1806.10155 [hep-ph]].
%62 citations counted in INSPIRE as of 05 Dec 2021

%\cite{Angelescu:2021lln}
\bibitem{Angelescu:2021lln}
A.~Angelescu, D.~Be\v{c}irevi\'c, D.~A.~Faroughy, F.~Jaffredo and O.~Sumensari,
%``Single leptoquark solutions to the B-physics anomalies,''
Phys. Rev. D \textbf{104} (2021) no.5, 055017
doi:10.1103/PhysRevD.104.055017
[arXiv:2103.12504 [hep-ph]].
%50 citations counted in INSPIRE as of 20 Dec 2021

%\cite{Greljo:2018tzh}
\bibitem{Greljo:2018tzh}
A.~Greljo, J.~Martin Camalich and J.~D.~Ruiz-\'Alvarez,
%``Mono-$\tau$ Signatures at the LHC Constrain Explanations of $B$-decay Anomalies,''
Phys. Rev. Lett. \textbf{122} (2019) no.13, 131803
doi:10.1103/PhysRevLett.122.131803
[arXiv:1811.07920 [hep-ph]].
%73 citations counted in INSPIRE as of 05 Dec 2021

%\cite{Husek:2021isa}
\bibitem{Husek:2021isa}
T.~Husek, K.~Monsalvez-Pozo and J.~Portoles,
%``Constraints on leptoquarks from lepton-flavour-violating tau-lepton processes,''
[arXiv:2111.06872 [hep-ph]].
%0 citations counted in INSPIRE as of 09 Dec 2021

%\cite{Haisch:2020xjd}
\bibitem{Haisch:2020xjd}
U.~Haisch and G.~Polesello,
%``Resonant third-generation leptoquark signatures at the Large Hadron Collider,''
JHEP \textbf{05} (2021), 057
doi:10.1007/JHEP05(2021)057
[arXiv:2012.11474 [hep-ph]].
%12 citations counted in INSPIRE as of 05 Jan 2022

%\cite{Aebischer:2018iyb}
\bibitem{Aebischer:2018iyb}
J.~Aebischer, J.~Kumar, P.~Stangl and D.~M.~Straub,
%``A Global Likelihood for Precision Constraints and Flavour Anomalies,''
Eur. Phys. J. C \textbf{79} (2019) no.6, 509
doi:10.1140/epjc/s10052-019-6977-z
[arXiv:1810.07698 [hep-ph]].
%94 citations counted in INSPIRE as of 20 Dec 2021

%\cite{Ferrara:1992yc}
\bibitem{Ferrara:1992yc}
S.~Ferrara, M.~Porrati and V.~L.~Telegdi,
%``$g=2$ as the natural value of the tree-level gyromagnetic ratio of elementary particles,''
Phys. Rev. D \textbf{46} (1992), 3529-3537
doi:10.1103/PhysRevD.46.3529
%178 citations counted in INSPIRE as of 14 Dec 2021

%\cite{Dorsner:2016wpm}
\bibitem{Dorsner:2016wpm}
I.~Dor\v{s}ner, S.~Fajfer, A.~Greljo, J.~F.~Kamenik and N.~Ko\v{s}nik,
%``Physics of leptoquarks in precision experiments and at particle colliders,''
Phys. Rept. \textbf{641} (2016), 1-68
doi:10.1016/j.physrep.2016.06.001
[arXiv:1603.04993 [hep-ph]].
%395 citations counted in INSPIRE as of 16 Dec 2021

%\cite{Zyla:2020zbs}
\bibitem{Zyla:2020zbs}
P.~A.~Zyla \textit{et al.} [Particle Data Group],
%``Review of Particle Physics,''
PTEP \textbf{2020} (2020) no.8, 083C01
doi:10.1093/ptep/ptaa104
%2512 citations counted in INSPIRE as of 22 Dec 2021

%\cite{Esteban:2020cvm}
\bibitem{Esteban:2020cvm}
I.~Esteban, M.~C.~Gonzalez-Garcia, M.~Maltoni, T.~Schwetz and A.~Zhou,
%``The fate of hints: updated global analysis of three-flavor neutrino oscillations,''
JHEP \textbf{09} (2020), 178
doi:10.1007/JHEP09(2020)178
[arXiv:2007.14792 [hep-ph]].
%330 citations counted in INSPIRE as of 21 Dec 2021

%\cite{King:2013eh}
\bibitem{King:2013eh}
S.~F.~King and C.~Luhn,
%``Neutrino Mass and Mixing with Discrete Symmetry,''
Rept. Prog. Phys. \textbf{76} (2013), 056201
doi:10.1088/0034-4885/76/5/056201
[arXiv:1301.1340 [hep-ph]].
%700 citations counted in INSPIRE as of 21 Dec 2021

%\cite{Lam:2007qc}
\bibitem{Lam:2007qc}
C.~S.~Lam,
%``Symmetry of Lepton Mixing,''
Phys. Lett. B \textbf{656} (2007), 193-198
doi:10.1016/j.physletb.2007.09.032
[arXiv:0708.3665 [hep-ph]].
%168 citations counted in INSPIRE as of 03 Dec 2021

%\cite{Hernandez:2012ra}
\bibitem{Hernandez:2012ra}
D.~Hernandez and A.~Y.~Smirnov,
%``Lepton mixing and discrete symmetries,''
Phys. Rev. D \textbf{86} (2012), 053014
doi:10.1103/PhysRevD.86.053014
[arXiv:1204.0445 [hep-ph]].
%201 citations counted in INSPIRE as of 06 Dec 2021

%\cite{Lam:2012ga}
\bibitem{Lam:2012ga}
C.~S.~Lam,
%``Finite Symmetry of Leptonic Mass Matrices,''
Phys. Rev. D \textbf{87} (2013) no.1, 013001
doi:10.1103/PhysRevD.87.013001
[arXiv:1208.5527 [hep-ph]].
%64 citations counted in INSPIRE as of 03 Dec 2021

%\cite{Holthausen:2012wt}
\bibitem{Holthausen:2012wt}
M.~Holthausen, K.~S.~Lim and M.~Lindner,
%``Lepton Mixing Patterns from a Scan of Finite Discrete Groups,''
Phys. Lett. B \textbf{721} (2013), 61-67
doi:10.1016/j.physletb.2013.02.047
[arXiv:1212.2411 [hep-ph]].
%97 citations counted in INSPIRE as of 03 Dec 2021

%\cite{King:2013vna}
\bibitem{King:2013vna}
S.~F.~King, T.~Neder and A.~J.~Stuart,
%``Lepton mixing predictions from $\Delta(6n^2)$ family Symmetry,''
Phys. Lett. B \textbf{726} (2013), 312-315
doi:10.1016/j.physletb.2013.08.052
[arXiv:1305.3200 [hep-ph]].
%102 citations counted in INSPIRE as of 05 Dec 2021

%\cite{Holthausen:2013vba}
\bibitem{Holthausen:2013vba}
M.~Holthausen and K.~S.~Lim,
%``Quark and Leptonic Mixing Patterns from the Breakdown of a Common Discrete Flavor Symmetry,''
Phys. Rev. D \textbf{88} (2013), 033018
doi:10.1103/PhysRevD.88.033018
[arXiv:1306.4356 [hep-ph]].
%62 citations counted in INSPIRE as of 03 Dec 2021

%\cite{Lavoura:2014kwa}
\bibitem{Lavoura:2014kwa}
L.~Lavoura and P.~O.~Ludl,
%``Residual $\mathbb{Z}_2 \times \mathbb{Z}_2$ symmetries and lepton mixing,''
Phys. Lett. B \textbf{731} (2014), 331-336
doi:10.1016/j.physletb.2014.03.001
[arXiv:1401.5036 [hep-ph]].
%27 citations counted in INSPIRE as of 03 Dec 2021

%\cite{Fonseca:2014koa}
\bibitem{Fonseca:2014koa}
R.~M.~Fonseca and W.~Grimus,
%``Classification of lepton mixing matrices from finite residual symmetries,''
JHEP \textbf{09} (2014), 033
doi:10.1007/JHEP09(2014)033
[arXiv:1405.3678 [hep-ph]].
%90 citations counted in INSPIRE as of 15 Dec 2021

%\cite{Joshipura:2014qaa}
\bibitem{Joshipura:2014qaa}
A.~S.~Joshipura and K.~M.~Patel,
%``Discrete flavor symmetries for degenerate solar neutrino pair and their predictions,''
Phys. Rev. D \textbf{90} (2014) no.3, 036005
doi:10.1103/PhysRevD.90.036005
[arXiv:1405.6106 [hep-ph]].
%21 citations counted in INSPIRE as of 03 Dec 2021

%\cite{Talbert:2014bda}
\bibitem{Talbert:2014bda}
J.~Talbert,
%``[Re]constructing Finite Flavour Groups: Horizontal Symmetry Scans from the Bottom-Up,''
JHEP \textbf{12} (2014), 058
doi:10.1007/JHEP12(2014)058
[arXiv:1409.7310 [hep-ph]].
%31 citations counted in INSPIRE as of 03 Dec 2021

%\cite{Yao:2015dwa}
\bibitem{Yao:2015dwa}
C.~Y.~Yao and G.~J.~Ding,
%``Lepton and Quark Mixing Patterns from Finite Flavor Symmetries,''
Phys. Rev. D \textbf{92} (2015) no.9, 096010
doi:10.1103/PhysRevD.92.096010
[arXiv:1505.03798 [hep-ph]].
%44 citations counted in INSPIRE as of 03 Dec 2021

%\cite{Lu:2016jit}
\bibitem{Lu:2016jit}
J.~N.~Lu and G.~J.~Ding,
%``Alternative Schemes of Predicting Lepton Mixing Parameters from Discrete Flavor and CP Symmetry,''
Phys. Rev. D \textbf{95} (2017) no.1, 015012
doi:10.1103/PhysRevD.95.015012
[arXiv:1610.05682 [hep-ph]].
%33 citations counted in INSPIRE as of 03 Dec 2021

%\cite{Varzielas:2016zuo}
\bibitem{Varzielas:2016zuo}
I.~de Medeiros Varzielas, R.~W.~Rasmussen and J.~Talbert,
%``Bottom-Up Discrete Symmetries for Cabibbo Mixing,''
Int. J. Mod. Phys. A \textbf{32} (2017) no.06n07, 1750047
doi:10.1142/S0217751X17500476
[arXiv:1605.03581 [hep-ph]].
%21 citations counted in INSPIRE as of 03 Dec 2021

%\cite{deMedeirosVarzielas:2019dyu}
\bibitem{deMedeirosVarzielas:2019dyu}
I.~de Medeiros Varzielas and J.~Talbert,
%``FCNC-free multi-Higgs-doublet models from broken family symmetries,''
Phys. Lett. B \textbf{800} (2020), 135091
doi:10.1016/j.physletb.2019.135091
[arXiv:1908.10979 [hep-ph]].
%3 citations counted in INSPIRE as of 03 Dec 2021

%\cite{Varzielas:2015iva}
\bibitem{Varzielas:2015iva}
I.~de Medeiros Varzielas and G.~Hiller,
%``Clues for flavor from rare lepton and quark decays,''
JHEP \textbf{06} (2015), 072
doi:10.1007/JHEP06(2015)072
[arXiv:1503.01084 [hep-ph]].
%185 citations counted in INSPIRE as of 10 Dec 2021

%\cite{Straub:2018kue}
\bibitem{Straub:2018kue}
D.~M.~Straub,
%``flavio: a Python package for flavour and precision phenomenology in the Standard Model and beyond,''
[arXiv:1810.08132 [hep-ph]].
%132 citations counted in INSPIRE as of 20 Dec 2021

%\cite{Aebischer:2018bkb}
\bibitem{Aebischer:2018bkb}
J.~Aebischer, J.~Kumar and D.~M.~Straub,
%``Wilson: a Python package for the running and matching of Wilson coefficients above and below the electroweak scale,''
Eur. Phys. J. C \textbf{78} (2018) no.12, 1026
doi:10.1140/epjc/s10052-018-6492-7
[arXiv:1804.05033 [hep-ph]].
%91 citations counted in INSPIRE as of 22 Dec 2021

%\cite{Efrati:2015eaa}
\bibitem{Efrati:2015eaa}
A.~Efrati, A.~Falkowski and Y.~Soreq,
%``Electroweak constraints on flavorful effective theories,''
JHEP \textbf{07} (2015), 018
doi:10.1007/JHEP07(2015)018
[arXiv:1503.07872 [hep-ph]].
%113 citations counted in INSPIRE as of 20 Dec 2021

%\cite{Falkowski:2017pss}
\bibitem{Falkowski:2017pss}
A.~Falkowski, M.~Gonz\'alez-Alonso and K.~Mimouni,
%``Compilation of low-energy constraints on 4-fermion operators in the SMEFT,''
JHEP \textbf{08} (2017), 123
doi:10.1007/JHEP08(2017)123
[arXiv:1706.03783 [hep-ph]].
%100 citations counted in INSPIRE as of 14 Dec 2021

%\cite{Altmannshofer:2014rta}
\bibitem{Altmannshofer:2014rta}
W.~Altmannshofer and D.~M.~Straub,
%``New physics in $b\rightarrow s$ transitions after LHC run 1,''
Eur. Phys. J. C \textbf{75} (2015) no.8, 382
doi:10.1140/epjc/s10052-015-3602-7
[arXiv:1411.3161 [hep-ph]].
%378 citations counted in INSPIRE as of 14 Dec 2021

%\cite{Descotes-Genon:2015uva}
\bibitem{Descotes-Genon:2015uva}
S.~Descotes-Genon, L.~Hofer, J.~Matias and J.~Virto,
%``Global analysis of $b\to s\ell\ell$ anomalies,''
JHEP \textbf{06} (2016), 092
doi:10.1007/JHEP06(2016)092
[arXiv:1510.04239 [hep-ph]].
%429 citations counted in INSPIRE as of 15 Dec 2021

%\cite{Jenkins:2013zja}
\bibitem{Jenkins:2013zja}
E.~E.~Jenkins, A.~V.~Manohar and M.~Trott,
%``Renormalization Group Evolution of the Standard Model Dimension Six Operators I: Formalism and lambda Dependence,''
JHEP \textbf{10} (2013), 087
doi:10.1007/JHEP10(2013)087
[arXiv:1308.2627 [hep-ph]].
%417 citations counted in INSPIRE as of 22 Dec 2021

%\cite{Jenkins:2013wua}
\bibitem{Jenkins:2013wua}
E.~E.~Jenkins, A.~V.~Manohar and M.~Trott,
%``Renormalization Group Evolution of the Standard Model Dimension Six Operators II: Yukawa Dependence,''
JHEP \textbf{01} (2014), 035
doi:10.1007/JHEP01(2014)035
[arXiv:1310.4838 [hep-ph]].
%410 citations counted in INSPIRE as of 22 Dec 2021

%\cite{Alonso:2013hga}
\bibitem{Alonso:2013hga}
R.~Alonso, E.~E.~Jenkins, A.~V.~Manohar and M.~Trott,
%``Renormalization Group Evolution of the Standard Model Dimension Six Operators III: Gauge Coupling Dependence and Phenomenology,''
JHEP \textbf{04} (2014), 159
doi:10.1007/JHEP04(2014)159
[arXiv:1312.2014 [hep-ph]].
%526 citations counted in INSPIRE as of 22 Dec 2021

%\cite{Aebischer:2015fzz}
\bibitem{Aebischer:2015fzz}
J.~Aebischer, A.~Crivellin, M.~Fael and C.~Greub,
%``Matching of gauge invariant dimension-six operators for $b\to s$ and $b\to c$ transitions,''
JHEP \textbf{05} (2016), 037
doi:10.1007/JHEP05(2016)037
[arXiv:1512.02830 [hep-ph]].
%105 citations counted in INSPIRE as of 03 Dec 2021

%\cite{Jenkins:2017jig}
\bibitem{Jenkins:2017jig}
E.~E.~Jenkins, A.~V.~Manohar and P.~Stoffer,
%``Low-Energy Effective Field Theory below the Electroweak Scale: Operators and Matching,''
JHEP \textbf{03} (2018), 016
doi:10.1007/JHEP03(2018)016
[arXiv:1709.04486 [hep-ph]].
%118 citations counted in INSPIRE as of 22 Dec 2021

%\cite{Aebischer:2017gaw}
\bibitem{Aebischer:2017gaw}
J.~Aebischer, M.~Fael, C.~Greub and J.~Virto,
%``B physics Beyond the Standard Model at One Loop: Complete Renormalization Group Evolution below the Electroweak Scale,''
JHEP \textbf{09} (2017), 158
doi:10.1007/JHEP09(2017)158
[arXiv:1704.06639 [hep-ph]].
%80 citations counted in INSPIRE as of 20 Dec 2021

%\cite{Jenkins:2017dyc}
\bibitem{Jenkins:2017dyc}
E.~E.~Jenkins, A.~V.~Manohar and P.~Stoffer,
%``Low-Energy Effective Field Theory below the Electroweak Scale: Anomalous Dimensions,''
JHEP \textbf{01} (2018), 084
doi:10.1007/JHEP01(2018)084
[arXiv:1711.05270 [hep-ph]].
%109 citations counted in INSPIRE as of 22 Dec 2021

%\cite{deBlas:2017xtg}
\bibitem{deBlas:2017xtg}
J.~de Blas, J.~C.~Criado, M.~Perez-Victoria and J.~Santiago,
%``Effective description of general extensions of the Standard Model: the complete tree-level dictionary,''
JHEP \textbf{03} (2018), 109
doi:10.1007/JHEP03(2018)109
[arXiv:1711.10391 [hep-ph]].
%114 citations counted in INSPIRE as of 22 Dec 2021

%\cite{Crivellin:2018yvo}
\bibitem{Crivellin:2018yvo}
A.~Crivellin, C.~Greub, D.~M\"uller and F.~Saturnino,
%``Importance of Loop Effects in Explaining the Accumulated Evidence for New Physics in B Decays with a Vector Leptoquark,''
Phys. Rev. Lett. \textbf{122} (2019) no.1, 011805
doi:10.1103/PhysRevLett.122.011805
[arXiv:1807.02068 [hep-ph]].
%116 citations counted in INSPIRE as of 20 Dec 2021

%\cite{Grzadkowski:2010es}
\bibitem{Grzadkowski:2010es}
B.~Grzadkowski, M.~Iskrzynski, M.~Misiak and J.~Rosiek,
%``Dimension-Six Terms in the Standard Model Lagrangian,''
JHEP \textbf{10} (2010), 085
doi:10.1007/JHEP10(2010)085
[arXiv:1008.4884 [hep-ph]].
%1532 citations counted in INSPIRE as of 22 Dec 2021

%\cite{Murgui:2019czp}
\bibitem{Murgui:2019czp}
C.~Murgui, A.~Pe\~nuelas, M.~Jung and A.~Pich,
%``Global fit to $b \to c \tau \nu$ transitions,''
JHEP \textbf{09} (2019), 103
doi:10.1007/JHEP09(2019)103
[arXiv:1904.09311 [hep-ph]].
%106 citations counted in INSPIRE as of 03 Dec 2021

%\cite{Shi:2019gxi}
\bibitem{Shi:2019gxi}
R.~X.~Shi, L.~S.~Geng, B.~Grinstein, S.~J\"ager and J.~Martin Camalich,
%``Revisiting the new-physics interpretation of the $b\to c\tau\nu$ data,''
JHEP \textbf{12} (2019), 065
doi:10.1007/JHEP12(2019)065
[arXiv:1905.08498 [hep-ph]].
%72 citations counted in INSPIRE as of 03 Dec 2021

%\cite{Blanke:2018yud}
\bibitem{Blanke:2018yud}
M.~Blanke, A.~Crivellin, S.~de Boer, T.~Kitahara, M.~Moscati, U.~Nierste and I.~Ni\v{s}and\v{z}i\'c,
%``Impact of polarization observables and $ B_c\to \tau \nu$ on new physics explanations of the $b\to c \tau \nu$ anomaly,''
Phys. Rev. D \textbf{99} (2019) no.7, 075006
doi:10.1103/PhysRevD.99.075006
[arXiv:1811.09603 [hep-ph]].
%83 citations counted in INSPIRE as of 03 Dec 2021

%\cite{Blanke:2019qrx}
\bibitem{Blanke:2019qrx}
M.~Blanke, A.~Crivellin, T.~Kitahara, M.~Moscati, U.~Nierste and I.~Ni\v{s}and\v{z}i\'c,
%``Addendum to \textquotedblleft{}Impact of polarization observables and $B_c\to \tau \nu$ on new physics explanations of the $b\to c \tau \nu$ anomaly'',''
doi:10.1103/PhysRevD.100.035035
[arXiv:1905.08253 [hep-ph]].
%78 citations counted in INSPIRE as of 03 Dec 2021

%\cite{Iguro:2020cpg}
\bibitem{Iguro:2020cpg}
S.~Iguro and R.~Watanabe,
%``Bayesian fit analysis to full distribution data of $ \overline{\mathrm{B}}\to {\mathrm{D}}^{\left(\ast \right)}\mathrm{\ell}\overline{\nu }:\left|{\mathrm{V}}_{\mathrm{cb}}\right| $ determination and new physics constraints,''
JHEP \textbf{08} (2020) no.08, 006
doi:10.1007/JHEP08(2020)006
[arXiv:2004.10208 [hep-ph]].
%21 citations counted in INSPIRE as of 05 Jan 2022

%\cite{Hiller:2018wbv}
\bibitem{Hiller:2018wbv}
G.~Hiller, D.~Loose and I.~Ni\v{s}and\v{z}i\'c,
%``Flavorful leptoquarks at hadron colliders,''
Phys. Rev. D \textbf{97} (2018) no.7, 075004
doi:10.1103/PhysRevD.97.075004
[arXiv:1801.09399 [hep-ph]].
%47 citations counted in INSPIRE as of 03 Dec 2021

%\cite{Hiller:2021pul}
\bibitem{Hiller:2021pul}
G.~Hiller, D.~Loose and I.~Ni\v{s}and\v{z}i\'c,
%``Flavorful leptoquarks at the LHC and beyond: spin 1,''
JHEP \textbf{06} (2021), 080
doi:10.1007/JHEP06(2021)080
[arXiv:2103.12724 [hep-ph]].
%21 citations counted in INSPIRE as of 13 Dec 2021

%\cite{Bordone:2019vic}
\bibitem{Bordone:2019vic}
M.~Bordone, M.~Jung and D.~van Dyk,
%``Theory determination of $\bar{B}\to D^{(*)}\ell^-\bar\nu$ form factors at $\mathcal{O}(1/m_c^2)$,''
Eur. Phys. J. C \textbf{80} (2020) no.2, 74
doi:10.1140/epjc/s10052-020-7616-4
[arXiv:1908.09398 [hep-ph]].
%55 citations counted in INSPIRE as of 20 Dec 2021

%\cite{ATLAS:2021jyv}
\bibitem{ATLAS:2021jyv}
G.~Aad \textit{et al.} [ATLAS],
%``Search for new phenomena in $pp$ collisions in final states with tau leptons, $b$-jets, and missing transverse momentum with the ATLAS detector,''
[arXiv:2108.07665 [hep-ex]].
%5 citations counted in INSPIRE as of 21 Dec 2021

%\cite{Drees:2013wra}
\bibitem{Drees:2013wra}
M.~Drees, H.~Dreiner, D.~Schmeier, J.~Tattersall and J.~S.~Kim,
%``CheckMATE: Confronting your Favourite New Physics Model with LHC Data,''
Comput. Phys. Commun. \textbf{187} (2015), 227-265
doi:10.1016/j.cpc.2014.10.018
[arXiv:1312.2591 [hep-ph]].
%289 citations counted in INSPIRE as of 03 Dec 2021

%\cite{Dercks:2016npn}
\bibitem{Dercks:2016npn}
D.~Dercks, N.~Desai, J.~S.~Kim, K.~Rolbiecki, J.~Tattersall and T.~Weber,
%``CheckMATE 2: From the model to the limit,''
Comput. Phys. Commun. \textbf{221} (2017), 383-418
doi:10.1016/j.cpc.2017.08.021
[arXiv:1611.09856 [hep-ph]].
%178 citations counted in INSPIRE as of 20 Dec 2021


%\cite{Alwall:2014hca}
\bibitem{Alwall:2014hca}
J.~Alwall, R.~Frederix, S.~Frixione, V.~Hirschi, F.~Maltoni, O.~Mattelaer, H.~S.~Shao, T.~Stelzer, P.~Torrielli and M.~Zaro,
%``The automated computation of tree-level and next-to-leading order differential cross sections, and their matching to parton shower simulations,''
JHEP \textbf{07}, 079 (2014)
doi:10.1007/JHEP07(2014)079
[arXiv:1405.0301 [hep-ph]].
%6636 citations counted in INSPIRE as of 11 May 2022

%\cite{Degrande:2011ua}
\bibitem{Degrande:2011ua}
C.~Degrande, C.~Duhr, B.~Fuks, D.~Grellscheid, O.~Mattelaer and T.~Reiter,
%``UFO - The Universal FeynRules Output,''
Comput. Phys. Commun. \textbf{183} (2012), 1201-1214
doi:10.1016/j.cpc.2012.01.022
[arXiv:1108.2040 [hep-ph]].
%1036 citations counted in INSPIRE as of 13 Jun 2022

%\cite{Alloul:2013bka}
\bibitem{Alloul:2013bka}
A.~Alloul, N.~D.~Christensen, C.~Degrande, C.~Duhr and B.~Fuks,
%``FeynRules  2.0 - A complete toolbox for tree-level phenomenology,''
Comput. Phys. Commun. \textbf{185} (2014), 2250-2300
doi:10.1016/j.cpc.2014.04.012
[arXiv:1310.1921 [hep-ph]].
%1937 citations counted in INSPIRE as of 15 Jun 2022

%\cite{NNPDF:2014otw}
\bibitem{NNPDF:2014otw}
R.~D.~Ball \textit{et al.} [NNPDF],
%``Parton distributions for the LHC Run II,''
JHEP \textbf{04} (2015), 040
doi:10.1007/JHEP04(2015)040
[arXiv:1410.8849 [hep-ph]].
%3093 citations counted in INSPIRE as of 16 Jun 2022

%\cite{Buckley:2014ana}
\bibitem{Buckley:2014ana}
A.~Buckley, J.~Ferrando, S.~Lloyd, K.~Nordstr\"om, B.~Page, M.~R\"ufenacht, M.~Sch\"onherr and G.~Watt,
%``LHAPDF6: parton density access in the LHC precision era,''
Eur. Phys. J. C \textbf{75} (2015), 132
doi:10.1140/epjc/s10052-015-3318-8
[arXiv:1412.7420 [hep-ph]].
%1189 citations counted in INSPIRE as of 14 Jun 2022


%\cite{Belanger:2021smw}
\bibitem{Belanger:2021smw}
G.~B\'elanger, A.~Bharucha, B.~Fuks, A.~Goudelis, J.~Heisig, A.~Jueid, A.~Lessa, K.~A.~Mohan, G.~Polesello and P.~Pani, \textit{et al.}
%``Leptoquark Manoeuvres in the Dark: a simultaneous solution of the dark matter problem and the $R_D$ anomalies,''
[arXiv:2111.08027 [hep-ph]].
%2 citations counted in INSPIRE as of 09 Dec 2021


%\cite{ATLAS:2019qpq}
\bibitem{ATLAS:2019qpq}
M.~Aaboud \textit{et al.} [ATLAS],
%``Searches for third-generation scalar leptoquarks in $\sqrt{s}$ = 13 TeV pp collisions with the ATLAS detector,''
JHEP \textbf{06} (2019), 144
doi:10.1007/JHEP06(2019)144
[arXiv:1902.08103 [hep-ex]].
%78 citations counted in INSPIRE as of 20 Dec 2021

%\cite{CMS:2018qqq}
\bibitem{CMS:2018qqq}
A.~M.~Sirunyan \textit{et al.} [CMS],
%``Constraints on models of scalar and vector leptoquarks decaying to a quark and a neutrino at $\sqrt{s}=$ 13 TeV,''
Phys. Rev. D \textbf{98} (2018) no.3, 032005
doi:10.1103/PhysRevD.98.032005
[arXiv:1805.10228 [hep-ex]].
%82 citations counted in INSPIRE as of 20 Dec 2021

%\cite{CMS:2018txo}
\bibitem{CMS:2018txo}
A.~M.~Sirunyan \textit{et al.} [CMS],
%``Search for a singly produced third-generation scalar leptoquark decaying to a $\tau$ lepton and a bottom quark in proton-proton collisions at $\sqrt{s} =$ 13 TeV,''
JHEP \textbf{07} (2018), 115
doi:10.1007/JHEP07(2018)115
[arXiv:1806.03472 [hep-ex]].
%55 citations counted in INSPIRE as of 19 Dec 2021

%\cite{CMS:2020wzx}
\bibitem{CMS:2020wzx}
A.~M.~Sirunyan \textit{et al.} [CMS],
%``Search for singly and pair-produced leptoquarks coupling to third-generation fermions in proton-proton collisions at s=13~TeV,''
Phys. Lett. B \textbf{819} (2021), 136446
doi:10.1016/j.physletb.2021.136446
[arXiv:2012.04178 [hep-ex]].
%29 citations counted in INSPIRE as of 20 Dec 2021

%\cite{Gripaios:2010hv}
\bibitem{Gripaios:2010hv}
B.~Gripaios, A.~Papaefstathiou, K.~Sakurai and B.~Webber,
%``Searching for third-generation composite leptoquarks at the LHC,''
JHEP \textbf{01} (2011), 156
doi:10.1007/JHEP01(2011)156
[arXiv:1010.3962 [hep-ph]].
%35 citations counted in INSPIRE as of 03 Dec 2021

%\cite{Brooijmans:2020yij}
\bibitem{Brooijmans:2020yij}
G.~Brooijmans, A.~Buckley, S.~Caron, A.~Falkowski, B.~Fuks, A.~Gilbert, W.~J.~Murray, M.~Nardecchia, J.~M.~No and R.~Torre, \textit{et al.}
%``Les Houches 2019 Physics at TeV Colliders: New Physics Working Group Report,''
[arXiv:2002.12220 [hep-ph]].
%44 citations counted in INSPIRE as of 11 Dec 2021

%\cite{ATLAS:2017mjy}
\bibitem{ATLAS:2017mjy}
M.~Aaboud \textit{et al.} [ATLAS],
%``Search for squarks and gluinos in final states with jets and missing transverse momentum using 36  fb$^{-1}$ of $\sqrt{s}=13$  TeV pp collision data with the ATLAS detector,''
Phys. Rev. D \textbf{97} (2018) no.11, 112001
doi:10.1103/PhysRevD.97.112001
[arXiv:1712.02332 [hep-ex]].
%225 citations counted in INSPIRE as of 15 Dec 2021

%\cite{ATLAS:2020syg}
\bibitem{ATLAS:2020syg}
G.~Aad \textit{et al.} [ATLAS],
%``Search for squarks and gluinos in final states with jets and missing transverse momentum using 139 fb$^{-1}$ of $\sqrt{s}$ =13 TeV $pp$ collision data with the ATLAS detector,''
JHEP \textbf{02} (2021), 143
doi:10.1007/JHEP02(2021)143
[arXiv:2010.14293 [hep-ex]].
%31 citations counted in INSPIRE as of 20 Dec 2021

%\cite{LHCb:2020lmf}
\bibitem{LHCb:2020lmf}
R.~Aaij \textit{et al.} [LHCb],
%``Measurement of $CP$-Averaged Observables in the $B^{0}\rightarrow K^{*0}\mu^{+}\mu^{-}$ Decay,''
Phys. Rev. Lett. \textbf{125} (2020) no.1, 011802
doi:10.1103/PhysRevLett.125.011802
[arXiv:2003.04831 [hep-ex]].
%149 citations counted in INSPIRE as of 03 Mar 2022

%\cite{LHCb:2020gog}
\bibitem{LHCb:2020gog}
R.~Aaij \textit{et al.} [LHCb],
%``Angular Analysis of the  $B^{+}\rightarrow K^{\ast+}\mu^{+}\mu^{-}$ Decay,''
Phys. Rev. Lett. \textbf{126} (2021) no.16, 161802
doi:10.1103/PhysRevLett.126.161802
[arXiv:2012.13241 [hep-ex]].
%62 citations counted in INSPIRE as of 03 Mar 2022
%\cite{LHCb:2021zwz}
\bibitem{LHCb:2021zwz}
R.~Aaij \textit{et al.} [LHCb],
%``Branching Fraction Measurements of the Rare $B^0_s\rightarrow\phi\mu^+\mu^-$ and $B^0_s\rightarrow f_2^\prime(1525)\mu^+\mu^-$- Decays,''
Phys. Rev. Lett. \textbf{127} (2021) no.15, 151801
doi:10.1103/PhysRevLett.127.151801
[arXiv:2105.14007 [hep-ex]].
%34 citations counted in INSPIRE as of 03 Mar 2022






%\cite{ATLAS:2019vcq}
\bibitem{ATLAS:2019vcq}
 [ATLAS],
%``Search for squarks and gluinos in final states with jets and missing transverse momentum using 139 fb$^{-1}$ of $\sqrt{s}$ =13 TeV $pp$ collision data with the ATLAS detector,''
ATLAS-CONF-2019-040.
%47 citations counted in INSPIRE as of 11 May 2022

%\cite{ATLAS:2019gti}
\bibitem{ATLAS:2019gti}
G.~Aad \textit{et al.} [ATLAS],
%``Search for direct stau production in events with two hadronic $\tau$-leptons in $\sqrt{s} = 13$ TeV $pp$ collisions with the ATLAS detector,''
Phys. Rev. D \textbf{101}, no.3, 032009 (2020)
doi:10.1103/PhysRevD.101.032009
[arXiv:1911.06660 [hep-ex]].
%58 citations counted in INSPIRE as of 11 May 2022



%\cite{deFavereau:2013fsa}
\bibitem{deFavereau:2013fsa}
J.~de Favereau \textit{et al.} [DELPHES 3],
%``DELPHES 3, A modular framework for fast simulation of a generic collider experiment,''
JHEP \textbf{02}, 057 (2014)
doi:10.1007/JHEP02(2014)057
[arXiv:1307.6346 [hep-ex]].
%2248 citations counted in INSPIRE as of 11 May 2022

%\cite{Dercks:2016npn}
%\bibitem{Dercks:2016npn}
%D.~Dercks, N.~Desai, J.~S.~Kim, K.~Rolbiecki, J.~Tattersall and T.~Weber,
%``CheckMATE 2: From the model to the limit,''
%Comput. Phys. Commun. \textbf{221}, 383-418 (2017)
%doi:10.1016/j.cpc.2017.08.021
%[arXiv:1611.09856 [hep-ph]].
%199 citations counted in INSPIRE as of 12 May 2022

%\cite{Cacciari:2008gp}
\bibitem{Cacciari:2008gp}
M.~Cacciari, G.~P.~Salam and G.~Soyez,
%``The anti-$k_t$ jet clustering algorithm,''
JHEP \textbf{04}, 063 (2008)
doi:10.1088/1126-6708/2008/04/063
[arXiv:0802.1189 [hep-ph]].
%8669 citations counted in INSPIRE as of 12 May 2022

%\cite{Cacciari:2011ma}
\bibitem{Cacciari:2011ma}
M.~Cacciari, G.~P.~Salam and G.~Soyez,
%``FastJet User Manual,''
Eur. Phys. J. C \textbf{72}, 1896 (2012)
doi:10.1140/epjc/s10052-012-1896-2
[arXiv:1111.6097 [hep-ph]].
%4728 citations counted in INSPIRE as of 12 May 2022

%\cite{Read:2002hq}
\bibitem{Read:2002hq}
A.~L.~Read,
%``Presentation of search results: The CL(s) technique,''
J. Phys. G \textbf{28}, 2693-2704 (2002)
doi:10.1088/0954-3899/28/10/313
%3674 citations counted in INSPIRE as of 12 May 2022

%\cite{Sjostrand:2014zea}
\bibitem{Sjostrand:2014zea}
T.~Sj\"ostrand, S.~Ask, J.~R.~Christiansen, R.~Corke, N.~Desai, P.~Ilten, S.~Mrenna, S.~Prestel, C.~O.~Rasmussen and P.~Z.~Skands,
%``An introduction to PYTHIA 8.2,''
Comput. Phys. Commun. \textbf{191}, 159-177 (2015)
doi:10.1016/j.cpc.2015.01.024
[arXiv:1410.3012 [hep-ph]].
%4308 citations counted in INSPIRE as of 12 May 2022
\end{thebibliography}
\end{document}